\documentclass[aps,pra,twocolumn,showpacs,amsmath,amssymb]{revtex4-1}
\usepackage{color}
\usepackage{graphicx}% Include figure files
\usepackage{dcolumn}% Align table columns on decimal point
\usepackage{bm}% bold math
\begin{document}

\title{Measures of electronic-vibrational entanglement and quantum coherence in a molecular system}

\author{Mihaela Vatasescu}
\email[]{mihaela\_vatasescu@yahoo.com}
%\thanks{}
\affiliation{Institute of Space Sciences - INFLPR,
MG-23, 77125 Bucharest-Magurele, Romania}

%\date{\today}

\begin{abstract}

We characterize both  entanglement and quantum coherence in a molecular
system  by connecting the linear entropy of electronic-nuclear
entanglement with Wigner-Yanase skew information measuring vibronic
coherence and local quantum uncertainty on electronic energy. 
Linear entropy of entanglement and quantifiers of quantum coherence
are derived for a molecular system described in a bipartite Hilbert space
$\cal{H}$$=$$\cal{H}$$_{el}$$\bigotimes$$\cal{H}$$_{vib}$ of finite dimension
$N_{el} \times N_v$, and relations between them are established.
For the specific case of the electronic-vibrational entanglement,
we find the linear entropy of entanglement  as having a more complex informational content
than the von Neumann entropy. By keeping the information carried by the vibronic coherences in a molecule,
linear entropy seizes vibrational motion in the electronic potentials as entanglement dynamics. 
We analyze entanglement oscillations in an isolated molecule,
and show examples for the control of entanglement dynamics in a
molecule through the creation of coherent vibrational wave packets in several electronic
potentials by  using chirped laser pulses.
\end{abstract}

%\pacs{03.67.Mn,34.80.Pa,33.80.Be,03.67.Bg,33.15.Vb}

\maketitle

\section{\label{sec:intro}Introduction}

Entanglement and coherence are both recognized as fundamental quantum
properties rooted in the superposition principle
\cite{4horodecki09,baumgratz14,streltsov-adesso15},
and as quantum resources \cite{baumgratz14,streltsov-adesso15,mhorodecki13,siewert14,brandao15}.
  Both are intertwined in two prominent research directions uniting
quantum information theory 
and molecular physics: quantum computation using molecular internal
 degrees of freedom \cite{lidar01,*lidar02,*tesch-riedle02,*palao-kosloff02,*vala02,*gollub06,*troppmann06,*brown06,*mishima08,*babikov14}
and quantum biology \cite{kbwhaley11,scholes12,kassal13,hildner13,scholes15}.
The first direction developed 
theoretical proposals for using coherent molecular superpositions to
implement quantum algorithms. In the second direction, the functional roles of entanglement and electronic
coherences in models of photosynthesis  are subject to an open debate
\cite{kbwhaley11,scholes12,popescu09,popescu12, scholes15}. Nevertheless, the considerable interest in the 
role played by quantum superpositions of electronic states in
photosynthetic light-harvesting complexes has flourished in femtosecond multidimensional
spectroscopy experiments revealing interesting coherence effects and
motivating advances in theory \cite{scholes12,hildner13,scholes15}.

Recently, entanglement and coherence were brought closer by
treating them  in the unified framework of resource theories
\cite{baumgratz14,girolami14,streltsov-adesso15,
mhorodecki13, brandao15}.
 The quantum theory of coherence 
being historically formulated in quantum optics \cite{glauber63,sudarshan63}, 
recent approaches have attempted to
develop a framework to quantify coherence in information theoretic
terms, following similar steps as for the theory of entanglement
\cite{baumgratz14,streltsov-adesso15}. 
In analogy with entanglement, coherence is now seen as a
quantum resource, and a quantitative theory of coherence was 
 formulated  as a resource theory \cite{baumgratz14,mhorodecki13,
   brandao15}.  Connections between entanglement and coherence are investigated, 
searching ``how can one resource emerge 
   quantitatively from the other''
   \cite{streltsov-adesso15}. It is interesting to underline that, unlike
entanglement and other resources in information theory, coherence is
basis-dependent \cite{scholes15,streltsov-adesso15}. Its meaning being given in
a reference basis of a particular observable, quantum coherence
appears as related to quantum uncertainty in a measurement of that
observable \cite{girolami13,girolami14}. Quantum correlations and
quantum uncertainty are hence brought together in 
 a context  enriched by the search for new relations among these fundamental quantum concepts. 

The present work searches for connections between electronic-vibrational
entanglement and quantum coherence in a molecular system. In a
previous paper  \cite{vatasescu2013} we have investigated the
entanglement between electronic and nuclear degrees of freedom created by vibronic
couplings which produce a pure entangled state in the bipartite
Hilbert space
$\cal{H}$$=$$\cal{H}$$_{el}$$\bigotimes$$\cal{H}$$_{vib}$. 
We have derived the von Neumann and linear entropies of
entanglement for the  $2 \times N_v$ and $3 \times N_v$ dimensions of
$\cal{H}$.
Here we derive the linear entropy of electronic-vibrational entanglement for a bipartite
Hilbert space with dimension $N_{el} \times N_v$, showing its
dependence on the vibronic coherences of the molecule. 
We show relations of electronic-nuclear
linear entropy of entanglement with several measures of coherence characterizing the bipartite
molecular system. We employ coherence quantifiers based on $l_1$ norm
\cite{baumgratz14} and Wigner-Yanase skew
information ${\cal I_S} (\rho, H)$ for a quantum state $\rho$ and
observable $H$ \cite{wigner63,girolami14}.

In a molecule with several populated electronic states, electronic and vibrational
degrees of freedom are entangled \cite{vatasescu2013}. Linear entropy of entanglement keeps the information
about the vibronic coherences existent in such a system, and shows an entanglement dynamics due to vibrational
motions in the electronic potentials. 
We analyze these entanglement oscillations in a molecule, considering the
temporal evolution of linear entropy after the action of laser pulses
which populate several electronic states. We show examples for the control of  entanglement dynamics
in a molecule by using chirped laser
pulses, whose parameters can be chosen to excite
various superpositions of vibrational states in each electronic
potential, allowing  specific quantum preparations and significant
changes in  entanglement dynamics.

The paper is structured as follows. Section \ref{sec:ElVibEnt}
outlines our model for entanglement in a pure state of the bipartite Hilbert
space $\cal{H}$$=$$\cal{H}$$_{el}$$\bigotimes$$\cal{H}$$_{vib}$. 
In Sec.~\ref{sec:NeuLinentr} we discuss
 the expressions for the von Neumann and linear entropies of
entanglement in a $2 \times N_v$ system, emphasizing the difference
between these two entanglement measures revealed 
 by their temporal behaviours in the case of an isolated molecule.
In Sec.~\ref{sec:linentrvibcoh} we derive
 the linear entropy of entanglement for an $N_{el} \times N_v$ system. Section \ref{sec:linentrosc}
analyzes the characteristic times of entanglement dynamics in an isolated molecule.
Section \ref{sec:EntCoh} characterizes quantum coherence in the pure entangled
  state $\hat{\rho}_{el,vib}(t)$, employing the resource approach, and
  shows the relation between the linear entropy of entanglement and 
the $l_1$ norm measure of coherence in the reduced electronic
state $\hat{\rho}_{el}(t)$. Section \ref{sec:CohUncert} connects
quantum coherence in the pure bipartite state $\hat{\rho}_{el,vib}(t)$
relative to the vibronic basis of the molecular Hamiltonian $\hat{H}_{mol}$, to quantum uncertainty in a
measurement of the observable $\hat{H}_{mol}$, 
and to the \text{"}velocity\text{"}
 of $\hat{\rho}_{el,vib}(t)$ evolution introduced by Anandan and Aharonov \cite{anandan90}.
In Sec.~\ref{sec:skewinf} are derived quantum coherence measures for
the bipartite system (el$\bigotimes$vib) based
on the Wigner-Yanase skew information, disclosing their connections
with the linear entropy of entanglement. Section \ref{sec:entdynvib}
contains examples showing entanglement oscillations in a molecule due
to vibronic coherences among 
 several electronic states populated by laser pulses. The control
of entanglement dynamics by using chirped laser pulses is shown in
the case of the $Cs_2$ molecule, for quantum preparations implying two
(Sec.~\ref{sec:3Su1g})
and three  (Sec.~\ref{sec:3su1g0g}) electronic states. 
 Conclusions are drawn in Sec.~\ref{conclu}.

\section{\label{sec:ElVibEnt} Entanglement in a pure state of the Hilbert space $\cal{H}$$=$$\cal{H}$$_{el}$$\bigotimes$$\cal{H}$$_{vib}$}

We consider  the
entanglement between electronic and vibrational degrees of freedom
created by vibronic couplings in a diatomic molecule described in the
Born-Oppenheimer (BO) approximation \cite{vatasescu2013}.  Neglecting the rotational degree of
freedom,  we focus on a pure entangled state $\hat{\rho}^2_{el,vib}=\hat{\rho}_{el,vib}$ of the Hilbert space
$\cal{H}$$=$$\cal{H}$$_{el}$$\bigotimes$$\cal{H}$$_{vib}$:
%
%Eq.~(\ref{psielvibket}) 
\begin{equation}
 \hat{\rho}_{el,vib}(t)=|\Psi_{el,vib}(t) > <\Psi_{el,vib}(t)|.
\label{densityop}
\end{equation}
$|\Psi_{el,vib}(t) >$  is an entangled state  of the bipartite system
 (el$\bigotimes$vib) created
by nonadiabatic couplings between BO molecular states (for example,
laser pulses coupling the electronic states),  having the form
\begin{equation}
|\Psi_{el,vib}(t) > = \sum_{\alpha=1}^{N_{el}} |\alpha> \bigotimes |\psi_{_\alpha}(t)>,
\label{psielvibket}
\end{equation}
where the summation is over the populated electronic channels $\alpha=\overline{1,N_{el}}$. 
The ket $|\Psi_{el,vib}(t) >$ denotes the molecular wavefunction
$\Psi_{el,vib}(\vec{r_{i}},R,t)$ which depends on the  electronic coordinates $\{ \vec{r_{i}} \}$ (expressed in
the molecule-fixed coordinate system), the internuclear distance $R$,
and the time $t$.
$|\alpha>$ denominates the electronic state
$\phi_{\alpha}^{el}(\vec{r_{i}};R)$,  and  $|\psi_{_\alpha}(t)>$  the
corresponding vibrational wave packet $\psi_{_\alpha}(R,t)$.
The electronic states $ |\alpha>=\phi_{\alpha}^{el}(\vec{r_{i}};R)$,
depending parametrically on R, are orthonormal eigenstates
of the electronic Hamiltonian $\hat{H}_{el}$, for which the "clamped nuclei"
electronic Schr\"odinger equation 
\begin{equation}
\hat{H}_{el} |\alpha>=V_{_\alpha}(R)|\alpha>
%\hat{H}_{el}\phi_{\alpha}^{el}(\vec{r_{i}};R)=V_{_\alpha}(R)\phi_{\alpha}^{el}(\vec{r_{i}};R)
\label{eqSeladiab}
\end{equation}
 gives the  adiabatic potential-energy surfaces $V_{_\alpha}(R)$
 as eigenvalues of $\hat{H}_{el}$ \cite{bookLefebField}.

The molecular Hamiltonian   is the sum of the electronic Hamiltonian $\hat{H}_{el}$ and the
  nuclear kinetic-energy $\hat{T}_R$:
\begin{equation}
\hat{H}_{mol}= \hat{H}_{el} + \hat{T}_R.
\end{equation}

Taking into account that in the BO approximation the nuclear motion in
an electronic state $|\alpha>$ 
is uniquely determined by the corresponding 
electronic potential  $V_{_\alpha}(R)$, the  Schr\"odinger equation giving the vibrational
eigenfunctions $\chi_{v_\alpha}(R)$ and vibrational energies
$E_{v_\alpha}$ is
\begin{equation}
[ \hat{T}_R   + V_{_\alpha}(R) ] |\chi_{v_\alpha}(R)> = E_{v_\alpha}
|\chi_{v_\alpha}(R)>.
\label{eqSvib}
\end{equation}
The eigenvectors
$\{|\chi_{v_\alpha}(R)>\}_{v_\alpha=\overline{1,N_\alpha}}$  form an orthonormal vibrational basis with dimension 
$N_\alpha$ corresponding to the electronic surface $\alpha$. The
vibrational wave packet corresponding to the electronic potential $\alpha$
can be developed in this basis as
$|\psi_{\alpha}(R,t)>=\sum_{v_\alpha=1}^{N_\alpha} c_{v_\alpha} (t)
|\chi_{v_\alpha}(R)> $,
with the complex coefficients $c_{v_\alpha} (t)$ providing the
probabilities $|c_{v_\alpha} (t)|^2$ for the population of the
vibrational states $|\chi_{v_\alpha}(R)>$.

Let us note that the product vectors $ |\alpha> |\chi_{v_\alpha}(R)>$
are eigenvectors of $\hat{H}_{mol}$:
\begin{equation}
[\hat{H}_{el} + \hat{T}_R] |\alpha> |\chi_{v_\alpha}(R)> = E_{v_\alpha} |\alpha> |\chi_{v_\alpha}(R)> .
\label{vibronicbasis}
\end{equation}
The product basis $\{|\alpha> |\chi_{v_\alpha}(R)>\}$ constitutes an orthonormal basis
set in $\cal{H}$$_{el}$$\bigotimes$$\cal{H}$$_{vib}$, and we shall
refer to it as the vibronic basis. We recall that
$\{|\alpha>\}$ constitutes a basis set for $\cal{H}$$_{el}$, but $\{ |\chi_{v_\alpha}(R)> \}$ is not a basis
 set for $\cal{H}$$_{vib}$, because vibrational functions corresponding to different electronic states 
are generally not orthogonal.

\subsection{\label{sec:NeuLinentr} Von Neumann and linear entropies of entanglement ($2 \times N_v$ system)}

We begin by discussing electronic-vibrational entanglement in the case of a bipartite Hilbert space 
$\cal{H}$$=$$\cal{H}$$_{el}$$\bigotimes$$\cal{H}$$_{vib}$ with
dimension  $2 \times N_v$. Denoting by
$|g>,|e>$ the two populated electronic states,  the bipartite pure
entangled state (\ref{psielvibket}) is
\begin{equation}
|\Psi_{el,vib}(t) > = |g> \bigotimes |\psi_{g}(R,t)>
+ |e> \bigotimes |\psi_{e}(R,t)>.
\label{2pure-elvib}
\end{equation}

In a previous work \cite{vatasescu2013} we have  analyzed   the entanglement between electronic and vibrational degrees of freedom in the bipartite pure state (\ref{2pure-elvib}) using two measures of entanglement: the von Neumann entropy and the linear entropy
of the reduced density operator $\hat{\rho}_{el}
=$Tr$_{vib}(\hat{\rho}_{el,vib})$. 

We have shown that for the state (\ref{2pure-elvib})  the  von Neumann entropy of
  entanglement  has a simple expression related to the populations of
the two electronic states $P_g(t)=<\psi_{g}(R,t)|\psi_{g}(R,t)>$,
$P_e(t)=<\psi_{e}(R,t)|\psi_{e}(R,t)>$ \cite{vatasescu2013}:
\begin{equation}
S_{vN}(\hat{\rho}_{el}(t)) = - P_g(t) \log_2 P_g(t)  -  P_e(t)\log_2 P_e(t). 
\label{vonNel}
\end{equation}
We have also derived the expression for the  linear entropy of entanglement, which is related to the purity of the reduced density operator of one of the two subsystems (we have considered 
 $\hat{\rho}_{el}$): 
\begin{equation}
L(t)=1- \text{Tr} (\hat{\rho}^2_{el}(t)).
\label{linentropy}
\end{equation}
With the normalization condition $P_g(t)+P_e(t)=1$,  the following expressions can be written for the purity and the
linear entropy \cite{vatasescu2013}:
\begin{equation}
\text{Tr} (\hat{\rho}^2_{el}(t)) = P^2_g(t) + P^2_e(t) + 2 |<\psi_{g}(R,t)|\psi_{e}(R,t)>|^2,
\label{purityred2}
\end{equation}
\begin{equation}
L(t) = 2P_g(t)P_e(t) - 2 |<\psi_{g}(R,t)|\psi_{e}(R,t)>|^2.
\label{linentr2}
\end{equation}
In Eq.~(\ref{linentr2}), $L(t)$ is bounded by $0 \le L(t) \le \frac{1}{2}$. Obviously, if only one of the electronic
states is populated, $S_{vN}(\hat{\rho}_{el}(t))$=0 and $L(t)=0$, and
the pure bipartite state is non-entangled.

Let us remark that,  in contrast to the von Neumann entropy
expressed by Eq.~(\ref{vonNel}),  the linear entropy of entanglement (Eq.~(\ref{linentr2}) ) depends not only on the populations of the
electronic states, but also on the overlap integral
$<\psi_{g}(R,t)|\psi_{e}(R,t)>$ of  the
vibrational wave packets belonging to the two electronic surfaces. In a molecule
this overlap integral is always
time evolving due to the vibrational motion. 
 Therefore, a remarkable difference between these two measures of 
 the molecular entanglement is revealed by their temporal behaviours in the case of an
isolated molecule. For an  isolated molecule,
the time evolution is
generated by the molecular Hamiltonian  $\hat{H}_{mol}$, which (without
introducing supplementary nonadiabatic radial couplings between the electronic states)
preserves constant population in
each electronic channel. Consequently, the von Neumann entropy of entanglement will
remain  constant, but
the linear entropy will show an entanglement
dynamics due to the vibrational motion in each electronic potential.
This entanglement dynamics illustrates the fact that, 
in a molecule with at least two electronic states populated
({\it i.e.} entanglement),
the electronic and nuclear degrees of freedom are not isolated one
from each other,  and the evolution directed by $\hat{H}_{mol}$
\footnote{Implying vibrational motions of the nuclear wave packets in
  the electronic states.}
 constitutes interaction between these two degrees of freedom,
{\it i.e.} a \text{"}nonlocal operation\text{"}  leading to entanglement dynamics.
Such a  temporal evolution of entanglement, due entirely to the
vibrational motion, without exchange of population between the
electronic channels, is \text{"}seen\text{"} by the linear entropy, but it is not
seized by the
von Neumann entropy of entanglement. 

 The difference shown here between these two entanglement measures  
could be considered as an example supporting the view that \text{"}different
 entanglement measures quantify different types of resources\text{"}
 \cite{siewert14}.  Nevertheless, in this specific case of molecular entanglement, the linear entropy of entanglement
 appears as a more complex informational quantity than the von Neumann
 entropy.  In this context it is interesting to recall the discussion about the 
 \text{"}conceptual inadequacy\text{"} of the von Neumann entropy in defining the
 information content of a quantum system, accompanied  by proposals for  a
new measure of the information content carried by the system, which has  proven to be
essentially the linear entropy \cite{zeilinger99, zeilinger01,luo06}.

\subsection{\label{sec:linentrvibcoh} Linear entropy of entanglement
  and  vibronic coherences ($N_{el} \times N_v$ system)}

For more than two electronic states, it is an intricate work to deduce
the von Neumann entropy of the reduced density matrix  $\hat{\rho}_{el}(t)$,
 but we can write the expression for the linear entropy of entanglement.
For $N_{el}$ populated electronic states of the molecule,
assuming a pure entangled state described by Eq.~(\ref{psielvibket}) in the bipartite Hilbert space of
dimension $N_{el} \times N_v$, the density operator (\ref{densityop}) can be written as 
\begin{equation}
 \hat{\rho}_{el,vib}(t)= \sum_{\alpha,\beta}^{N_{el}}  | \alpha><\beta
 | \bigotimes  |\Psi_{\alpha}(t) > <\Psi_{\beta}(t)|,
\label{densityopN}
\end{equation}
and the reduced electronic density operator $\hat{\rho}_{el}=$Tr$_{vib}(\hat{\rho}_{el,vib})$$=\sum_{j=1}^{N_v}<j|\hat{\rho}_{el,vib}|j>$ (with $\{ |j > \}_{j=1,N_v}$ a complete orthonormal 
basis of $\cal{H}$$_{vib}$) becomes
\begin{equation}
 \hat{\rho}_{el}(t)= \sum_{\alpha,\beta}^{N_{el}}  | \alpha ><\beta |  <\Psi_{\beta}(R,t) |\Psi_{\alpha}(R,t)>.
\label{densityopelN}
\end{equation}
Therefore, one obtains for  the purity of the reduced electronic density
\begin{equation}
\text{Tr}_{el} (\hat{\rho}^2_{el}(t)) = \sum_{\alpha,\beta}^{N_{el}} |<\psi_{_\alpha}(R,t)|\psi_{_\beta}(R,t)>|^2.
\label{purityred}
\end{equation}
Taking into account the normalization condition
$\sum_{\alpha=1}^{N_{el}}P_{_\alpha}(t)=1$ for the total
population, with $P_{_\alpha}(t)$$=$$<\psi_{\alpha}(R,t)|\psi_{\alpha}(R,t)>$,
the linear entropy $L(t)=1-\text{Tr}_{el}(\hat{\rho}^2_{el}(t))$ can be
written as
\begin{equation}
   L(t) = 2 \sum_{\alpha,\beta, \alpha \ne \beta}^{N_{el}} [ P_{_\alpha}(t) P_{_\beta}(t) -
   |<\psi_{\alpha}(R,t)|\psi_{\beta}(R,t)>|^2 ].
   \label{linentrgen}
\end{equation}
The linear entropy defined by  Eq.~(\ref{linentrgen}) is bounded by $0
\le L(t) \le 1-\frac{1}{N_{el}}$, which shows the increasing of
$L(t)$ maximum by increasing the number of populated electronic
states $N_{el}$.

The linear entropy (\ref{linentrgen}) is related to
the vibronic coherences of the molecular system.
The connection appears through the matrix elements of the density
operator $\hat{\rho}_{el,vib}(t)$ in the vibronic basis $\{|\alpha>
|\chi_{v_\alpha}(R)>\}$, constituted by the eigenvectors of $\hat{H}_{mol}= \hat{H}_{el} + \hat{T}_R$.

The entangled state (\ref{psielvibket}) can be written as
\begin{equation}
|\Psi_{el,vib}(t) > = \sum_{\alpha=1}^{N_{el}} |\alpha> \bigotimes 
\sum_{v_{\alpha}=1}^{N_{\alpha}} c_{v_{\alpha}} (t) |\chi_{v_{\alpha}}(R)>,
\label{psielvibcv}
\end{equation}
where each nuclear wave packet $|\psi_{\alpha}(R,t)>$  was developed
 in the corresponding vibrational basis
$\{ |\chi_{v_{\alpha}}(R)> \}_{v_{\alpha}=\overline{1,N_{\alpha}}}$. The dimension of the 
vibrational Hilbert space $\cal{H}$$_{vib}$ is $N_v=\sum_{\alpha=1}^{N_{el}} N_{\alpha}$.
The complex coefficients $c_{v_{\alpha}} (t)$ give the population
probabilities $|c_{v_\alpha} (t)|^2$ for the 
vibrational levels  $\{v_{\alpha}\}$, and
 the population of an electronic state $\alpha$ is
$P_{\alpha}= \sum_{v_{\alpha}=1}^{N_{\alpha}} |c_{v_{\alpha}} (t)|^2$.

The populations and coherences \cite{cohen} of the molecular system are obtained 
as matrix elements of the density operator $\hat{\rho}_{el,vib}(t)$:
\begin{equation}
\rho_{\alpha v_{\alpha},\beta v_{\beta}}(t)=
<\alpha| <\chi_{v_{\alpha}}| \hat{\rho}_{el,vib}(t)|\chi_{v_{\beta}}> |\beta> 
= c_{v_{\alpha}} (t)c^{*}_{v_{\beta}} (t).
\label{coherences}
\end{equation}
The diagonal matrix elements $\rho_{\alpha v_{\alpha}, \alpha v_{\alpha}}(t)= |c_{v_ {\alpha}} (t)|^2$
are the  vibrational populations,
and the off-diagonal matrix elements (\ref{coherences})
 give the vibronic coherences (for $\alpha$$\ne$$\beta$),
as well as the vibrational coherences
$\rho_ {\alpha v_{\alpha}, \alpha v'_{\alpha}}(t)=$$c_{v_{\alpha}} (t)c^{*}_{v'_{\alpha}}(t)$.

Using  Eq.~(\ref{psielvibcv}) to rewrite  Eq.~(\ref{linentrgen}), it
appears that, besides the electronic
 populations $P_{_\alpha}(t)$,  the linear entropy contains explicitely the vibronic coherences
 $\rho_{\beta v_{\beta},\alpha v_{\alpha}}(t)$$=c^{*}_{v_{\alpha}} (t) c_{v_{\beta}} (t)$ 
 modulated by the overlap integral $<\chi_{v_ {\alpha}}
 (R)|\chi_{v_{\beta}}(R)>$ of the
vibrational wave functions:
\begin{eqnarray}
   L(t) = 2 \sum_{\alpha,\beta, \alpha \ne \beta}^{N_{el}} [ P_{_\alpha}(t) P_{_\beta}(t)\nonumber\\
-| \sum_{v_{\alpha}=1}^{N_{\alpha}} \sum_{v_{\beta}=1}^{N_{\beta}}       
c^{*}_{v_{\alpha}} (t) c_{v_{\beta}} (t)  <\chi_{v_{\alpha}} (R)|\chi_{v_{\beta}}(R)>|^2 ]
   \label{linentrgencoh}
\end{eqnarray}

Linear entropy dependence on the vibronic
coherences is a key property, which connects this entanglement measure
with coherence quantifiers in a molecule, as we will show in the next sections.
It is also due to this property that vibrational motion in
at least  two electronic states is seized  as giving a  dynamics of
entanglement between electronic and vibrational degrees of freedom.

\subsection{\label{sec:linentrosc} Linear entropy dynamics due to vibrational motions in the
  electronic potentials: Entanglement oscillations in an isolated molecule.}

In Sec.~\ref{sec:NeuLinentr}  we have shown  that, in contrast to the
von Neumann entropy of entanglement,  the linear entropy
``understands'' the vibrational motion in the electronic potentials
as entanglement dynamics.  Sec.~\ref{sec:linentrvibcoh} has developed
further this observation, showing that linear entropy keeps the
information carried by the vibronic coherences of the molecular
system. This section will specify  the characteristic
times of entanglement dynamics due to vibrational motion.

In a previous work \cite{vatasescu2013} we have analyzed the electronic- vibrational
entanglement dynamics produced by laser pulses coupling 
electronic states, focusing on the dynamics during  pulses. 
Here we will closely look at entanglement dynamics after a laser pulse (or a pulse sequence)
populates  several electronic states. The time evolution after
pulses is determined  by the molecular Hamiltonian $H_{mol}$, and in
the absence of other  nonadiabatic radial couplings which could transfer
population between the electronic channels, the electronic populations
will remain constant. In this case, as it is shown in  Sec.~\ref{sec:NeuLinentr},
the von Neumann entropy of entanglement remains constant too, but the
linear entropy shows an entanglement dynamics due to the dependence on
the vibronic coherences among electronic channels. This entanglement
dynamics entirely due to the vibrational motion in the electronic
channels of an ``isolated molecule''  will be analyzed in
this section. Numerical examples will be shown in the last section of
this paper.

Let us consider an isolated molecule with at least two populated electronic
states, whose time evolution generated by
$\hat{H}_{mol}$ leaves these electronic populations constant in
time. The linear entropy of entanglement is
expressed by Eq.~(\ref{linentrgen}), and we look at its time evolution
due to vibrational motion.
We begin by noting the two extreme cases of zero and maximal overlap
between vibrational wave packets. i) For nonoverlapping
vibrational wave packets, $<\psi_{\alpha}(R,t)|\psi_{\beta}(R,t)>=0$, 
 $L(t)$ will remain constant in time if the electronic populations are constant.
ii)  In principle a separability could appear even if several electronic
surfaces are populated, if the vibrational wave packets corresponding
to different electronic surfaces are
very similar both in R and in t. We can see that if $|\psi_{\alpha}(R,t)> \approx |\psi_{\beta}(R,t)>$, $L(t) \to
0$, and the entanglement is absent. Obviously this is a very
particular case, which would be possible in a special configuration
of electronic potentials with similar shapes. 

Returning to the general case, let us see  the characteristic
times appearing in $L(t)$ evolution due to vibrational motion.
Taking into account that the electronic channels $\alpha$ are not
coupled, the time evolution of each vibrational wave packet
$|\psi_{\alpha}(R,t)>=\sum_{v_\alpha=1}^{N_\alpha} c_{v_\alpha} (t) |\chi_{v_\alpha}(R)>  $  in the electronic potential $V_{\alpha}(R)$
 is directed by the Schr\"odinger equation $[\hat{T}_R +V_{\alpha}(R)] |\psi_{\alpha}(R,t)> =
 i\hbar{\partial }/ {\partial t} |\psi_{\alpha}(R,t)>$.  The probability
 amplitudes $c_{v_\alpha} (t)$ have the simple form:
\begin{equation}
 c_{v_\alpha} (t)=c_{v_\alpha}(t_i) e^{-\frac{i}{\hbar} E_{v_\alpha}(t-t_i)},
\end{equation}
where $t_i$ is a time moment after which the electronic channels can be considered
 uncoupled, and $E_{v_\alpha}$ is the vibrational energy corresponding to the
 vibrational function  $|\chi_{v_\alpha}(R)>$ (see Eq.~(\ref{eqSvib})).

We shall take the example of two electronic channels, for which the
linear entropy is given by Eq.~(\ref{linentr2}).
 If the populations $P_g, P_e$ rest constant in time for $t \ge t_i$,
 with $P_g=P_g(t_i)$ and
$P_e=P_e(t_i)$,  the time evolution of the linear entropy in
Eq. (\ref{linentr2}) is given by the term
\begin{widetext}
\begin{eqnarray}
 |<\psi_{g}(R,t)|\psi_{e}(R,t)>|^2 = %\nonumber \\
\sum_{v_g=1}^{N_g} \sum_{v'_g=1}^{N_g}\sum_{v_e=1}^{N_e} \sum_{v'_e=1}^{N_e}
c^{*}_{v_g} (t_i) c_{v_e} (t_i)  c_{v'_g} (t_i) c^{*}_{v'_e} (t_i) \nonumber \\
<\chi_{v_g} (R)|\chi_{v_e}(R)><\chi_{v'_e} (R)|\chi_{v'_g}(R)>
e^{\frac{i}{\hbar} [(E_{v_g}- E_{v'_g})  -  (E_{v_e}- E_{v'_e}) ] (t-t_i)}.
\label{oscil}
\end{eqnarray}
\end{widetext}
Therefore, the time evolution of $L(t)$   will show oscillations with the
characteristic times:
\begin{equation}
T_{osc}=\frac{2 \pi \hbar}{\Delta E_{v_gv'_gv_ev'_e}},
\end{equation}
with $\Delta E_{v_gv'_gv_ev'_e}=|(E_{v_g}- E_{v'_g})  -  (E_{v_e}- E_{v'_e})|$. Depending on the vibrational levels
populated in each electronic surface, the oscillation periods contributing in the time evolution are determined
by energy intervals varying from $\Delta E_{v_gv'_gv_ev'_e}=||E_{v_g}- E_{v'_g}| -  |E_{v_e}- E_{v'_e}||$
to $\Delta E_{v_gv'_gv_ev'_e}=|E_{v_g}- E_{v'_g}| +  |E_{v_e}-
E_{v'_e}|$.  On the other hand, the oscillations will have amplitudes depending on the 
populations of the vibrational levels (through the coefficients $c_{v} (t_i)$) and on the
vibrational overlaps. 

Let us specify two particular simple cases:

$\bullet$ In a $2 \times 2$ system, with
one vibrational level in each electronic state, the linear entropy does not vary in time:
$L_{v_g v_e}(t) = 2 |c_{v_g}(t)|^2 |c_{v_e}(t)|^2  (1-|<\chi_{v_g}|\chi_{v_e}> |^2)$.

$\bullet$ In a $2 \times 3$ system, supposing one level $v_g$
populated in the electronic state $g$, and two levels
$v_e, v'_e$ in the  electronic state $e$ , $L(t)$ will show oscillations
given by $\cos[(E_{v_e}- E_{v'_e})(t-t_i)/\hbar]$, with a
characteristic time $T_{osc}=2 \pi
\hbar/|E_{v_e}- E_{v'_e}|$. If $v_e, v'_e$ are neighboring levels, this time is the
vibrational period of  $v_e$, $T_{osc}=T_{vib}(v_e)$.

An interesting question is how large the time variations of the linear
entropy can be, during the time evolution under $H_{mol}$. Obviously
the dynamics of the electronic-nuclear entanglement  depends on the
electronic potentials of the molecule and on the specific quantum
preparations.
Therefore, for a particular molecule, the entanglement dynamics
can be directed by laser pulses able to excite vibrational
superpositions in several electronic states, creating a molecule with
\text{''}multiple vibrations\text{''}.  
In Sec.~\ref{sec:entdynvib} we will expose examples showing the control of entanglement dynamics
in a molecule with laser pulses coupling electronic states.

\section{\label{sec:EntCoh} Quantum Coherence in the pure entangled
  state $\hat{\rho}_{el,vib}(t)$}

The entangled state $|\Psi_{el,vib}(t)>$   (Eq.~(\ref{psielvibket}))
may be regarded as a superposition of eigenstates of $\hat{H}_{mol}$,
and therefore can also be characterized as a coherent state. The concept of \text{"}state
coherence\text{"} \cite{scholes15} refers  to  a superposition of 
eigenstates of an operator and  implies a basis-dependent coherence definition \cite{cohen,scholes15}.
In the present case, one
may speak of coherence relative to the  vibronic
basis, but also of coherence relative to a local vibrational basis
(related to a specific electronic state). 
If only one electronic state is populated, $|\Psi_{el,vib}(t) >$ being constituted by a superposition of vibrational states of this electronic state,
obviously $\hat{\rho}_{el,vib}(t)$ is not anymore an entangled state,
but it may  still be a coherent state,
due to the presence of vibrational coherences.

We will explore the connections between entanglement and coherence
in the state $|\Psi_{el,vib}(t)>$,  showing that  linear entropy of
entanglement is connected to measures of coherence in the
molecular system.

\subsection{\label{sec:CohResource} Coherence  in the framework of resource theories}

A variety of measures are used to characterize coherence,  generally
being functions of the density matrix' off-diagonal elements in a
reference basis. Recently, Baumgratz {\it et al} \cite{baumgratz14}
proposed to use the framework of resource theories 
\cite{mhorodecki13,brandao15} for the
quantification of coherence in information theoretic terms,
 following the approach previously established for entanglement.
In the resource approach, the quantification of  coherence
begins with the characterization of the \text{"}incoherent
states\text{"} (having a basis dependent definition: a state is incoherent in a
particular basis if its density matrix is diagonal in this basis)  and
of the corresponding class of \text{"}incoherent operations\text{"} (\text{"}free\text{"} operations that do not create coherence
from an incoherent state) \cite{baumgratz14}. A set of conditions  a  proper measure
of coherence should satisfy is proposed,
in analogy with well known requirements
from entanglement theory, such as the basic conditions
of monotonicity under incoherent operations and of the coherence
quantifier becoming zero for all
incoherent states. Several coherence
quantifiers satisfying these conditions are discussed in
Ref. \cite{baumgratz14}, such as the $l_1$ norm, the relative entropy
of coherence, and coherence quantifiers based on distance measures.

We will make two observations in order to connect the case treated here  to the coherence approach
formulated  in Ref. \cite{baumgratz14}, based on the identification of
 incoherent states and incoherent operations.

i)  The pure entangled state $\hat{\rho}_{el,vib}(t)$ is
a bipartite coherent state in the vibronic basis.  A question of
interest is the following: Is it
possible to found a basis in which this density matrix would become
diagonal, defining an incoherent state in that basis ? The
answer is no, there is no basis in the bipartite Hilbert space in which 
the entangled state $\hat{\rho}_{el,vib}(t)$ would become
incoherent. It can be shown that this requirement would imply identical
vibrational wave packets (up to a constant complex factor) in all 
 electronic states, which supposes a factorization dissolving the entanglement.
On the other hand, it can be shown that  bipartite incoherent states are always
separable \cite{streltsov-adesso15}, while $\hat{\rho}_{el,vib}(t)$ is
an entangled state.

ii) Temporal evolution generated by  $\hat{H}_{mol}$
constitutes an \text{"}incoherent operation\text{"}.  In Ref.~\cite{streltsov-adesso15} 
it is shown  that entanglement can be generated from coherent states via
incoherent operations,  which introduces  an interrogation about the \text{"}maximization of the output
entanglement\text{"}.
For an isolated molecule, it is  $\hat{H}_{mol}$ that generates the
evolution of the coherent entangled state $\hat{\rho}_{el,vib}(t)$
(Eq.~(\ref{evolHmol})).  We have already shown that
temporal evolution under  $\hat{H}_{mol}$
 creates an entanglement dynamics, and consequently a maximization or a
minimization of entanglement. In the last section we will show
specific examples of temporal evolution in a molecule illustrating 
  significant linear entropy variations during time evolution.

Unlike entanglement,  coherence is basis-dependent
\cite{streltsov-adesso15}.  Here we shall refer to two reference
bases for molecular coherence.  We shall discuss coherence of
the bipartite state $\hat{\rho}_{el,vib}(t)$  relative to the vibronic basis
$\{|\alpha>|\chi_{v_\alpha}(R)>\}$,  and coherence of the electronic
state $\hat{\rho}_{el}(t)$ taking the basis  $\{|\alpha>\}$ of the
electronic adiabatic states as reference basis.

We begin by using  the $l_1$ norm,  defined as \cite{baumgratz14}
\begin{equation}
C_{l_1}(\hat{\rho})=\sum_{i,j,i \ne j} |\rho_{ij}|
\label{l1norm}
\end{equation}
as a coherence quantifier.  For simplicity, we consider the $2 \times
N_v$ case, the two electronic states being $|g>,|e> $.
$C_{l_1}(\hat{\rho}_{el,vib} )$ is a measure for the coherence of the pure state
$\hat{\rho}_{el,vib}(t)$ in the vibronic basis, and for the  $2 \times
N_v$ case is
\begin{eqnarray}
C_{l_1}(\hat{\rho}_{el,vib} )=2 \{ \sum_{v_g=1}^{N_g}\sum_{v_e=1}^{N_e} | c_{v_g} (t)c^{*}_{v_e}(t)|\nonumber\\
+ \sum_{v_g,v'_g, v_g \ne v'_g }^{N_g} | c_{v_g} (t)c^{*}_{v'_g}(t)| 
+ \sum_{v_e,v'_e, v_e \ne v'_e }^{N_e} | c_{v_e} (t)c^{*}_{v'_e}(t)| \}.\nonumber\\
\end{eqnarray}
The first term is a measure of the vibronic
coherence, the others being quantifiers of vibrational coherence in each
electronic state.  As a measure of coherence
in the  global pure entangled state, $C_{l_1}(\hat{\rho}_{el,vib} )$  remains constant in time for an isolated molecule. 

Let us also consider the coherence of the reduced electronic state $\hat{\rho}_{el}(t)$ in the
electronic adiabatic basis $\{|g>,|e>\}$, measured by $C_{l_1}(
\hat{\rho}_{el})$. Taking into account the definition (\ref{l1norm})
and Eq.~(\ref{densityopelN}), we find
\begin{equation}
C_{l_1}( \hat{\rho}_{el}) = 2 |<\psi_{g}(R,t)|\psi_{e}(R,t)>|,
\label{cl1el}
\end{equation}
and then the following relation to the linear entropy of entanglement:
\begin{equation}
L(t) = 2P_g(t)P_e(t) - \frac{1}{2} [C_{l_1}( \hat{\rho}_{el})]^2.
\label{linentr2l1norm}
\end{equation}
Eq.~(\ref{linentr2l1norm}) constitutes a first relation established
here between a measure  of  entanglement in the bipartite molecular system
and a measure of coherence for the electronic subsystem.
The measure $C_{l_1}( \hat{\rho}_{el})$ of the electronic coherence
varies in time for an isolated molecule in the bipartite pure state $\hat{\rho}_{el,vib}(t)$,
being a sensor of quantum correlations in this entangled state. The temporal variation of $L(t)$ due
to vibrational motions reflects the time variation of coherence
of the reduced electronic state $\hat{\rho}_{el}(t)$. 
When the  overlap $|<\psi_{g}(R,t)|\psi_{e}(R,t)>|$ is large,
$C_{l_1}( \hat{\rho}_{el})$ is large, and $L(t)$
diminishes. Intuitively, a large overlap indicates the same
spatial localization of the vibrational wave packets, favoring the
separability between electronic  and vibrational degrees of
freedom, and consequently diminishing the entanglement.

\subsection{\label{sec:CohUncert} Quantum coherence, quantum
  uncertainty in energy, and the \text{"}velocity\text{"}
 of $\hat{\rho}_{el,vib}(t)$ evolution}

Quantum coherence has been shown to be closely related to quantum
uncertainty in a measurement \cite{girolami13,girolami14}. For the
system treated in this paper, the connection between quantum coherence
and quantum uncertainty could be formulated in the following manner:
$\hat{\rho}_{el,vib}(t)$ shows coherence in $\hat{H}_{mol}$ basis
because $\hat{\rho}_{el,vib}(t)$ does not commute with $\hat{H}_{mol}$
\footnote{Being neither an eigenstate of $\hat{H}_{mol}$,
nor a mixture of eigenstates of $\hat{H}_{mol}$, but a superposition of
eigenstates of $\hat{H}_{mol}$},
and therefore a quantum measurement
of the observable $\hat{H}_{mol}$ in the state $\hat{\rho}_{el,vib}(t)$ is
characterized by a quantum uncertainty due to quantum
coherence. Indeed, the commutator 
\begin{eqnarray}
[\hat{H}_{mol}, \hat{\rho}_{el,vib}(t)]= \nonumber\\
\sum_{\alpha,\beta} \sum_{v_{\alpha},v_{\beta}}
 c_{v_{\alpha}}(t)c^{*}_{v_{\beta}}(t) (E_{v_\alpha} - E_{v_\beta}) |\alpha><\beta|
|\chi_{v_{\alpha}}>< \chi_{v_{\beta}}| \nonumber\\
\end{eqnarray}
is nonzero due to  nonzero coherences of
$\hat{\rho}_{el,vib}(t)$,  and it determines the time evolution of the
density operator if  $\hat{H}_{mol}$ is the Hamiltonian generating the evolution of the
system:
\begin{equation}
i \hbar \frac{d \hat{\rho}_{el,vib}(t)}{dt} = [\hat{H}_{mol},
\hat{\rho}_{el,vib}(t)].
\label{evolHmol}
\end{equation}

For the pure state $|\Psi_{el,vib}(t) >$, the energy uncertainty on an
outcome associated with a  measurement of $\hat{H}_{mol}$ is exclusively due to the
quantum coherence \cite{girolami14}, being  measured by the
energy variance ${\cal V}(\hat{H}_{mol}, |\Psi_{el,vib}(t) >)$  ({\it i.e.} the mean square deviation from the average value,
$(\Delta \hat{H}_{mol})^2=<\hat{H}^2_{mol}>- < \hat{H}_{mol}>^2$):
\begin{eqnarray}
(\Delta \hat{H}_{mol})^2={\cal V}(\hat{H}_{mol}, |\Psi_{el,vib}(t) >) \nonumber\\
= \frac{1}{2} \sum_{\alpha,\beta} \sum_{v_{\alpha},v_{\beta}}
(E_{v_\beta} - E_{v_\alpha})^2 |c_{v_{\alpha}}(t)|^2 |c_{v_{\beta}}(t)|^2
\label{varianceHmol}
\end{eqnarray}
Anandan and Aharonov \cite{anandan90} have given a \text{"}geometric meaning to
the uncertainty in energy\text{"} for a quantum system, connecting the
energy uncertainty to the \text{"}distance along the evolution of the system\text{"}
in the projective Hilbert space.
For a pure state,  the uncertainty
in energy  gives the squared \text{"}velocity\text{"} of the
state evolution \cite{anandan90,brody11}. 
Here the equation illustrating this idea is
\begin{eqnarray}
 \text{Tr}_{el,vib} \left[ \frac{d \hat{\rho}_{el,vib}(t)}{dt} \frac{d
   \hat{\rho}_{el,vib}(t)}{dt} \right] = \frac{2}{ \hbar^2}(\Delta
 \hat{H}_{mol})^2.
\label{velostate}
\end{eqnarray}
Eq.~(\ref{velostate})  recovers a relation for the pure states evolution
appearing in Ref. \cite{brody11}, being connected to a time-energy
uncertainty relation deduced in quantum state estimation theory.

\section{\label{sec:skewinf} Wigner-Yanase skew information as a
  measure of quantum coherence and uncertainty in energy measurement. 
Connection with
linear entropy of entanglement.}

In Ref.~\cite{girolami14}, Girolami proposed a quantum coherence measure based
on the Wigner-Yanase skew information, satisfying the criteria enounced in
Ref. \cite{baumgratz14} which treats coherence in the framework of the
quantum information theory. Central to this approach is the observation
that  quantum uncertainty in measuring an observable $K$ in a
state $\rho$ is due to coherence shown by $\rho$ in $K$ eigenbasis.
 
The skew information was introduced by Wigner and Yanase as a measure
for the information content of a quantum state $\rho$ not commuting
with (skew to) an observable $K$ \cite{wigner63}:
\begin{equation}
{\cal I_S} (\rho,K)= -\frac{1}{2} \text{Tr} [\sqrt{\rho},K]^2.
\label{skewinf}
\end{equation}
Wigner and Yanase have shown that ${\cal I_S}$ satisfies the
requirements of an information measure \cite{wigner63}, relevant to
the measurement of observables which do not commute with a conserved
additive quantity $K$. The skew information is positive and vanishes
only if the state $\rho$ and observable $K$ commute. ${\cal
  I_S}(\rho,K)$ is always smaller than the variance of  $K$, ${\cal
  I_S}(\rho,K) \le {\cal V}(\rho,K)$,  and equals  the variance  
for a pure state $\rho=|\psi><\psi|=\sqrt{\rho}$. 

The skew information is a well known information-theoretic quantity,
associated with the quantum
Fisher information \cite{luo-prl03,luo-ams03},  quantum correlations
\cite{chen05,luo-oh12,girolami13}, and uncertainty relations \cite{luo-prl03,luo05,luo06,furuichi10}.
We refer to \cite{luo-oh12} for several related interpretations of  ${\cal I_S}$.
The skew information (Eq.~\ref{skewinf}) depends on both the state
$\rho$ and the observable $K$, 
being a measure of the quantum uncertainty
of $K$ in the state $\rho$ \cite{luo05,luo06,furuichi10,girolami13}, 
and a measure of the  $K$ coherence of the state $\rho$ \cite{girolami14}.

Here we employ the skew information as a measure of quantum
coherence and quantum uncertainty in the pure entangled state
$\hat{\rho}_{el,vib}(t)$  and in the reduced electronic state 
 $\hat{\rho}_{el}$,  taking as observables  the Hamiltonians  $\hat{H}_{mol}$ or $\hat{H}_{el}$. 
Considering coherence in the case of the bipartite entangled state
$\hat{\rho}_{el,vib}(t)$, as well as for the reduced electronic state 
 $\hat{\rho}_{el}$, we will provide links between entanglement
 and coherence measures.

We calculate the skew information  in the bipartite  state
$\hat{\rho}_{el,vib}(t)$ for the observables $\hat{H}_{mol}$ and
$\hat{H}_{el}\bigotimes \hat{I}_v$, 
as well as the skew information in the reduced electronic state 
 $\hat{\rho}_{el}$ for the electronic Hamiltonian $\hat{H}_{el}$. 

Eq.~(\ref{skewinf}) is usually rewritten as \cite{wigner63}
\begin{equation}
{\cal I_S} (\rho,H)=  Tr ({\rho},H^2) - Tr (\sqrt{\rho} H \sqrt{\rho} H),
\label{skewinf2}
\end{equation}
where we have considered as observable a Hamiltonian $H$. In an
orthonormal basis $\{|u_n>\}$ of $H$ (with eigenvalues $E_n$ and
eigenvectors $|u_n>$, $H |u_n> = E_n |u_n>$), Eq.~(\ref{skewinf2})
becomes  \cite{luo-ams03}:
\begin{equation}
{\cal I_S} (\rho,H)= \frac{1}{2} \sum_{m,n} (E_m - E_n)^2 |<u_m| \sqrt{\rho} |u_n>|^2.
\label{skewinfH}
\end{equation}
Eq.~(\ref{skewinfH}) will be used to obtain skew information relative to the molecular system. 
For the pure bipartite state $\hat{\rho}_{el,vib}(t)$, using the
vibronic basis of $\hat{H}_{mol}$ (Eq.~(\ref{vibronicbasis})), one obtains
\begin{eqnarray}
{\cal I_S} (\hat{\rho}_{el,vib}(t), \hat{H}_{mol}) = {\cal V}(\hat{H}_{mol}, |\Psi_{el,vib}(t) >) \nonumber\\
=
\frac{1}{2} \sum_{\alpha,\beta} \sum_{v_{\alpha},v_{\beta}}
(E_{v_\beta} - E_{v_\alpha})^2 |c_{v_{\alpha}}(t)|^2 |c_{v_{\beta}}(t)|^2.
\label{sinfHmol}
\end{eqnarray}

Eqs.~(\ref{sinfHmol}) and (\ref{varianceHmol}) express the same result, taking into account that
for a pure state $\rho=\sqrt{\rho}$. ${\cal I_S} [\hat{\rho}_{el,vib}(t), \hat{H}_{mol}]$ represents a measure
of the coherence of $\hat{\rho}_{el,vib}(t)$ relative to the vibronic basis of $\hat{H}_{mol}$, and a measure
of the quantum uncertainty on a measurement pertaining to $\hat{H}_{mol}$ in the state $\hat{\rho}_{el,vib}(t)$.
We recall also the original meaning of ${\cal I_S}$ \cite{wigner63} as
information content of $\hat{\rho}_{el,vib}(t)$ on the values of
observables not commuting with $\hat{H}_{mol}$.

We will show that the linear entropy of entanglement
(Eqs.~(\ref{linentr2}) and  (\ref{linentrgen})) is
related to the skew information for the observable $\hat{H}_{el}$.
For this end, we  compute ${\cal I_S} (\hat{\rho}_{el}, \hat{H}_{el})$ and ${\cal I_S}
(\hat{\rho}_{el,vib}, \hat{H}_{el}\bigotimes \hat{I}_v)$.  Both are
connected to the measurement of the local observable
$\hat{H}_{el}$ in the correlated quantum systems  (el$\bigotimes$vib).
We shall treat separately the $2 \times N_v$ and $N_{el} \times N_v$ cases.

\subsection{\label{sec:skewinf2Nv} Wigner-Yanase skew information for the electronic
  Hamiltonian $\hat{H}_{el}$, in the quantum states $\hat{\rho}_{el}$ and $\hat{\rho}_{el,vib}$
($2 \times N_v$ case)}

\subsubsection{$\mathbf {\cal I_S} (\hat{\rho}_{el}, \hat{H}_{el})$}

The skew information
\begin{equation}
{\cal I_S}(\hat{\rho}_{el}, \hat{H}_{el})   = -\frac{1}{2}
\text{Tr}_{el}[\sqrt{\hat{\rho}_{el}},\hat{H}_{el}]^2 
\end{equation}
for the local state $\hat{\rho}_{el}$ with respect
to the local observable $\hat{H}_{el}$
has several related interpretations: as a measure of the noncommutativity between $\hat{\rho}_{el}$  and
$\hat{H}_{el}$; as information content of $\hat{\rho}_{el}$ with respect to
$\hat{H}_{el}$, and with respect to observables not commuting with $\hat{H}_{el}$;
as a measure of quantum uncertainty on $\hat{H}_{el}$ in the state $\hat{\rho}_{el}$;
and as a measure of the $\hat{H}_{el}$ coherence in the state $\hat{\rho}_{el}$.
Moreover, ${\cal  I_S}(\hat{\rho}_{el}, \hat{H}_{el})$ is  a quantity
with  information content 
on a local observable ($\hat{H}_{el}$)
of a quantum subsystem ($\hat{\rho}_{el}$), and therefore it will also
 keep the trace of
quantum correlations in the  bipartite system $\hat{\rho}_{el,vib}$.

We have employed Eq.~(\ref{skewinfH}) to obtain ${\cal  I_S}(\hat{\rho}_{el}, \hat{H}_{el})$, taking into
account that  the electronic states  $\{|g>,|e>\}$ form an orthonormal basis for
$\hat{H}_{el}$, with eigenvalues  $V_g(R)$, $V_e(R)$ (the adiabatic electronic potentials):
\begin{eqnarray}
\hat{H}_{el}|g>=V_g(R)|g>~,~\hat{H}_{el}|e>=V_e(R)|e>.
\end{eqnarray}
The matrix of the reduced electronic density $\hat{\rho}_{el}=\text{Tr}_{vib}[\hat{\rho}_{el,vib}]=\sum_{j=1}^{N_v}<j|\hat{\rho}_{el,vib}|j>$ ( with  $ \{ |j > \}_{j=1,N_v}$ a complete orthonormal 
basis of $\cal{H}$$_{vib}$) in the electronic basis
$\{|g>,|e>\}$ is
\begin{eqnarray}
( \hat{\rho}_{el} )_{ \{ g,e \} } 
=
\left(\begin{array}{cc}
   P_g &  <\psi_{e}|\psi_{g}>\\
  <\psi_{g}|\psi_{e}> & P_e
 \end{array} \right).
\label{matdensel2}
\end{eqnarray}

Let us observe that in the  $\{|g>,|e>\}$  basis the commutator between
$\hat{\rho}_{el}$ and $ \hat{H}_{el}$ is
\begin{eqnarray}
( [ \hat{\rho}_{el},  \hat{H}_{el} ] )_{ \{ g,e \} }= \nonumber\\
\left(\begin{array}{cc}
   0      &   (V_e- V_g) <\psi_{e}|\psi_{g}> \\
 (V_g- V_e) <\psi_{g}|\psi_{e}>  &  0
 \end{array} \right), \nonumber\\
\label{comutrohel}
\end{eqnarray}
 and, with Eq.~(\ref{skewinfH}), the skew information ${\cal I_S} (\hat{\rho}_{el}, \hat{H}_{el})$
in this basis becomes
\begin{equation}
{\cal I_S} (\hat{\rho}_{el}, \hat{H}_{el}) =  [V_g(R) - V_e(R)]^2 
\frac{|<\psi_{g}(R,t)|\psi_{e}(R,t)>|^2}{ 1+ \sqrt{2L(t)} }.
\label{skewel2}
\end{equation}
Eq.~(\ref{skewel2})   shows that  ${\cal I_S}(\hat{\rho}_{el},\hat{H}_{el})$ 
has a time evolution determined by the vibronic coherences (see 
Eq.~(\ref{oscil})) and the linear entropy of entanglement $L(t)$, having the following relation 
 to the
$l_1$ norm measure of coherence $C_{l_1}( \hat{\rho}_{el}(t))$:
\begin{equation}
 {\cal I_S} (\hat{\rho}_{el}, \hat{H}_{el}) =  [V_g(R) - V_e(R)]^2 
\frac{ [ C_{l_1}(\hat{\rho}_{el}]^2}{4 [1+ \sqrt{2L(t)}] }  
\end{equation}
${\cal I_S} (\hat{\rho}_{el}, \hat{H}_{el})$ depends on the internuclear distance $R$ and the time $t$.
It indicates how the uncertainty related to a measurement of the
electronic energy  in the electronic
subsystem depends on the difference between the electronic potentials at  particular $R$, and on the
time evolutions of the coherence and entanglement.
 ${\cal I_S}(\hat{\rho}_{el}, \hat{H}_{el})$
may be considered as a quantifier of quantum uncertainty on
$\hat{H}_{el}$ in the state $\hat{\rho}_{el}(t)$.

\subsubsection{$\mathbf {\cal I_S} (\hat{\rho}_{el,vib},
  \hat{H}_{el}\bigotimes \hat{I}_v)$} 

 The skew information ${\cal I_S}
 (\hat{\rho}_{el,vib},\hat{H}_{el}\bigotimes \hat{I}_v)$ 
 (with $\hat{I}_v$ the identity operator in the vibrational Hilbert
 space $\cal{H}$$_{vib}$)  reflects the concept of \text{"}local quantum uncertainty\text{"} introduced in
Ref.~\cite{girolami13}, being associated to the measurement of local
observables in correlated quantum systems \footnote{Ref.~\cite{girolami13}
shows that the \text{"}local quantum uncertainty\text{"} is a measure
of bipartite quantum correlations and it is an entanglement monotone
for a pure bipartite state $\hat{\rho}$.}.

Taking  $\{|g>,|e>\}$ as the electronic basis for
$\hat{H}_{el}$, with eigenvalues  $V_g(R)$, $V_e(R)$, 
 the matrix of the density operator  $\hat{\rho}_{el,vib}$ in this basis is
\begin{eqnarray}
&&\left( \hat{\rho}_{el,vib} \right)_{ \{ g,e \} }=
\left(\begin{array}{lc}
   |\psi_{g}><\psi_{g}| &  |\psi_{g}><\psi_{e}|\\
  |\psi_{e}><\psi_{g}| & |\psi_{e}><\psi_{e}|
 \end{array} \right),
\label{matdens2}
\end{eqnarray}
and the commutator between
$\hat{\rho}_{el,vib}$ and $ \hat{H}_{el}\bigotimes \hat{I}_v$ is given by
\begin{eqnarray}
( [ \hat{\rho}_{el,vib},  \hat{H}_{el}\bigotimes \hat{I}_v ] )_{ \{ g,e \} } = \nonumber\\
\left(\begin{array}{cc}
   0      &   (V_e- V_g) |\psi_{g}><\psi_{e}| \\
 (V_g- V_e) |\psi_{e}><\psi_{g}|  &  0
 \end{array} \right).
\label{comutrohelvib}
\end{eqnarray}
The skew information can be expressed as
\begin{eqnarray}
{\cal I_S} (\hat{\rho}_{el,vib}, \hat{H}_{el}\bigotimes \hat{I}_v) 
=-\frac{1}{2}
\text{Tr}_{el,vib}[\sqrt{\hat{\rho}_{el,vib}},\hat{H}_{el}\bigotimes \hat{I}_v]^2 \nonumber\\
= 
\sum_{j=1}^{N_v} <j |  
\frac{1}{2} \sum_{m,n} (E_m - E_n)^2 |<u_m| \sqrt{\hat{\rho}_{el,vib}} |u_n>|^2
|j>, \nonumber\\
\label{skewinfj}
\end{eqnarray}
where $ \{ |j > \}_{j=1,N_v}$ is a complete orthonormal basis in
$\cal{H}$$_{vib}$, and $\{|u_n>\}$ an orthonormal basis of $\hat{H}_{el}$ (with eigenvalues $E_n$,
$\hat{H}_{el} |u_n> = E_n |u_n>$).  Therefore, we obtain
\begin{equation}
{\cal I_S} (\hat{\rho}_{el,vib}, \hat{H}_{el}\bigotimes \hat{I}_v) = [V_g(R) - V_e(R)]^2 P_g (t) P_e (t).
\label{skewelvib2}
\end{equation}
The skew information (\ref{skewelvib2})
is a measure of quantum uncertainty on
a measurement of the local observable $\hat{H}_{el}$ (electronic energy) in the bipartite state
$\hat{\rho}_{el,vib}(t)$. As $\hat{\rho}_{el,vib}(t)$ is the state of a bipartite entangled system, and
$\hat{H}_{el}\bigotimes \hat{I}_v$  a local observable, ${\cal I_S} (\hat{\rho}_{el,vib},
\hat{H}_{el}\bigotimes \hat{I}_v)$ may be considered as a witness of the bipartite
quantum correlations. 

\subsubsection{Connection with $L(t)$}

Now we can see that the linear entropy of entanglement $L(t)$ given by
Eq.~(\ref{linentr2}) has an interesting connection
with the two types of skew information corresponding to the electronic Hamiltonian:
\begin{eqnarray}
{\cal I_S} (\hat{\rho}_{el,vib}, \hat{H}_{el}\bigotimes \hat{I}_v) -  [1+ \sqrt{2L(t)}]
 {\cal I_S} (\hat{\rho}_{el}, \hat{H}_{el})  \nonumber \\ 
 = [V_g(R) - V_e(R)]^2  \frac{L(t)}{2}. 
\label{skewlinentropy}
\end{eqnarray}

The relation (\ref{skewlinentropy}) can be seen as expressing the
quantum correlations in the bipartite system $\hat{\rho}_{el,vib}(t)$
from the \text{"}perspective of the local observable\text{"} $\hat{H}_{el}$  (see also Ref. \cite{luo-oh12}).

\subsection{Wigner-Yanase skew information for the electronic
  Hamiltonian $\hat{H}_{el}$ in the $ N_{el} \times N_v$ case}

We shall now deduce the skew information ${\cal I_S} (\hat{\rho}_{el}, \hat{H}_{el})$ and ${\cal I_S}
(\hat{\rho}_{el,vib}, \hat{H}_{el}\bigotimes \hat{I}_v)$   for the general case of  $N_{el}$ populated
electronic states, for which the
 density operators $\hat{\rho}_{el,vib}(t)$  and $\hat{\rho}_{el}(t)$
 are expressed in Eqs.~(\ref{densityopN}) and (\ref{densityopelN}). 
The skew information can be obtained  in the adiabatic basis
$\{|\alpha>\}$ of the electronic Hamiltonian $H_{el}$, having the 
  adiabatic potential-energy surfaces $V_{_\alpha}(R)$ as eigenvalues
(Eq.~(\ref{eqSeladiab})). In
the electronic basis  $ \{ |\alpha > \}_{j=1,N_{el}}$ 
the density operators have the  matrix elements
\begin{eqnarray}
 <\alpha | \hat{\rho}_{el,vib} |\beta > = | \psi_{_\alpha}> < \psi_{_\beta}| ,\\
 <\alpha | \hat{\rho}_{el} |\beta > = < \psi_{_\beta} | \psi_{_\alpha}>.
\end{eqnarray}
Using Eqs.~(\ref{skewinfH})  and ~(\ref{skewinfj}) we obtain
\begin{equation}
{\cal I_S} (\hat{\rho}_{el,vib}, \hat{H}_{el}\bigotimes \hat{I}_v) =
\sum_{\alpha,\beta, \alpha \ne \beta}^{N_{el}}[V_{\alpha}(R) - V_{\beta}(R)]^2 P_{\alpha} (t) P_{\beta}(t),
\label{skewelvibN}
\end{equation}
\begin{equation}
{\cal I_S} (\hat{\rho}_{el}, \hat{H}_{el}) =  \sum_{\alpha,\beta, \alpha \ne \beta}^{N_{el}}[V_{\alpha}(R) - V_{\beta}(R)]^2   | <\alpha |\sqrt{\hat{\rho}_{el}}| \beta >|^2.
\label{skewelN}
\end{equation}
Therefore, it appears that for more than two electronic states, the quantum
correlations become more intricate, and the relation between the skew
information and the linear entropy of entanglement
is not as simple as in Eq.~(\ref{skewlinentropy}).
We observe that the difference
${\cal I_S}(\hat{\rho}_{el,vib},\hat{H}_{el}\bigotimes \hat{I}_v)$$-{\cal I_S}(\hat{\rho}_{el}, \hat{H}_{el})$  
is a sum containing  correlations terms
of the type $[ P_{\alpha} (t) P_{\beta}(t)-$$| <\alpha |\sqrt{\hat{\rho}_{el}}| \beta >|^2 ]$
as significant quantities, whereas  the linear
entropy $L(t)$ expressed in Eq.~(\ref{linentrgen}) is a sum containing  terms 
 $[ P_{_\alpha}(t) P_{_\beta}(t)
 -$$|<\psi_{\alpha}(R,t)|\psi_{\beta}(R,t)>|^2 ]$.

Let us also observe  that the coherence measures
$C_{l_1}(\hat{\rho}_{el})$ and  ${\cal I_S} (\hat{\rho}_{el}, \hat{H}_{el})$,  pertaining to 
the reduced electronic system, contain  the
quantities $|<\psi_{\alpha}(R,t)|\psi_{\beta}(R,t)>| $ related to the vibronic coherences, as we have shown in
Sec.~\ref{sec:linentrvibcoh}.  Therefore, like the linear entropy of
entanglement $L(t)$, these coherence measures reflect the bipartite
correlations and are varying in time due to the vibrational motion.

\section{\label{sec:entdynvib} Entanglement oscillations in a molecule
with  several populated electronic states}

The aim of this section is to show examples of electronic-nuclear entanglement dynamics in a
molecule, after the action of laser pulses, which  populate several
electronic states. We have shown that linear entropy of
entanglement has a time evolution due to the vibronic coherences arisen in the molecular
system, being connected to coherence measures analyzed in the
previous section.  We will give examples of  entanglement and
coherence dynamics, in a molecule with
 two or three electronic states populated by
chirped laser pulses. The purpose is double: on the one hand, to show
the entanglement oscillations due to vibrational motions in
realistic electronic potentials of a molecule, and to have an insight
about the amplitude of $L(t)$ variations over time;
 on the other hand, to show the control of the  entanglement dynamics
 by using chirped laser
pulses, whose parameters can be chosen to excite
various superpositions of vibrational states in each electronic
potential.  Specific quantum preparations according to the 
 shapes of the electronic curves lead to various possibilities of
entanglement control in a given molecule.

\begin{figure}
\includegraphics[width=0.95\columnwidth]{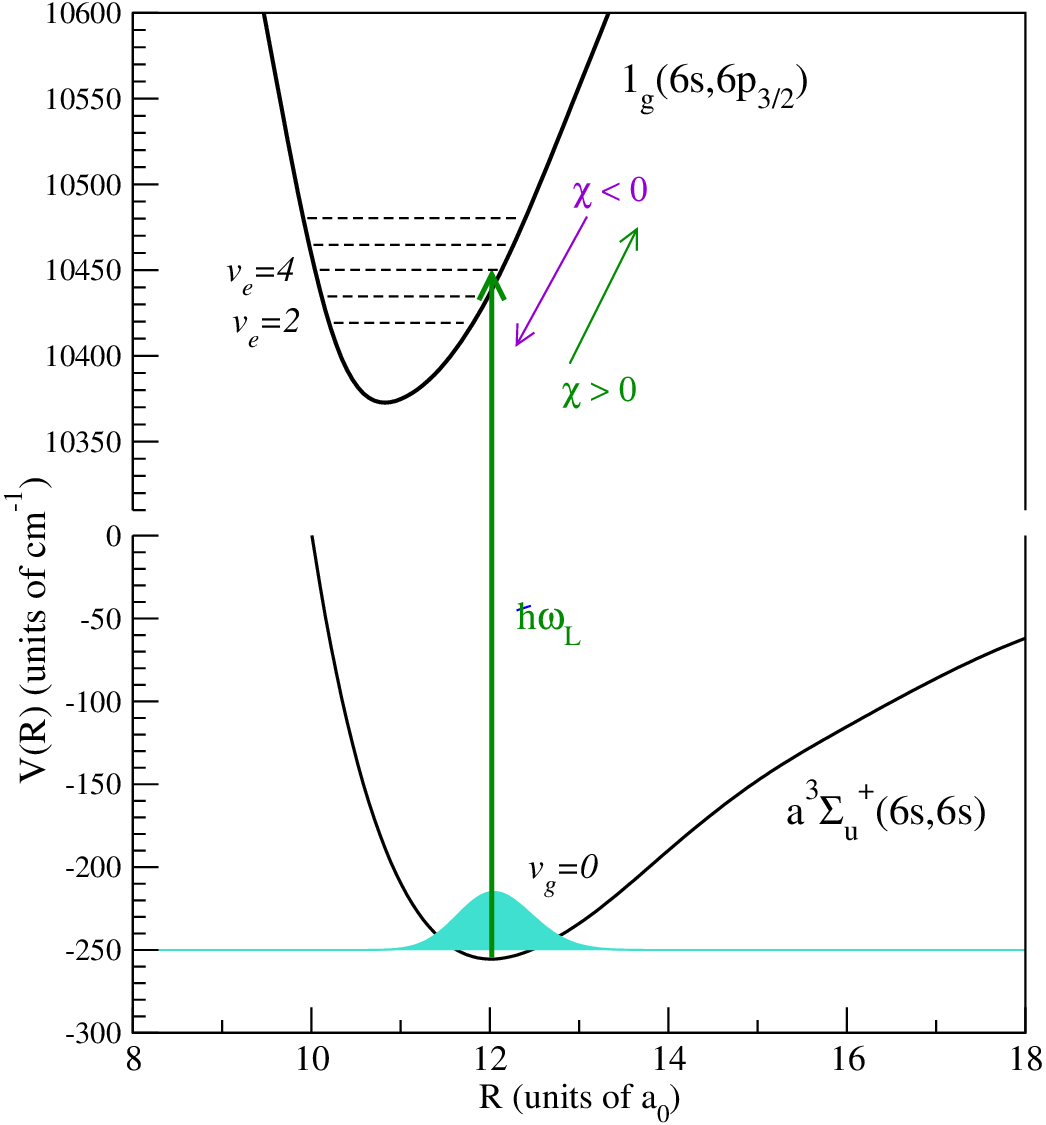}%
 \caption{\label{fig1_3Su1g} (Color online) $a^3\Sigma_{u}^{+}(6s,6s)$ and $1_g(6s,6p_{3/2})$ electronic
potentials of Cs$_2$, coupled by a chirped laser pulse with central
energy $\hbar \omega_L=$ 10695 cm$^{-1}$.
The initial state of the process is the vibrational wavefunction with $v_g=0$ of the $a^3\Sigma_{u}^{+}(6s,6s)$
electronic state. The pulse excites several vibrational levels $v_e$
in the $1_g(6s,6p_{3/2})$ electronic
potential. The energy origin is taken to be the
dissociation limit $E_{6s+6s}=0$ of the $a^3\Sigma_{u}^{+}(6s,6s)$
potential.}
\end{figure}

We will take as examples transitions implying the electronic states
$a^3\Sigma_{u}^{+}(6s,6s)$, $1_g(6s,6p_{3/2})$, and
$0^-_g(6s,6p_{3/2})$ of the Cs$_2$ molecule.
Sec.~\ref{sec:3Su1g} contains a paradigmatic example of two electronic
states coupled by a chirped laser pulse which transfers population
from the ground electronic state to several vibrational levels of the
excited state. We will show that, depending on the quantum preparation, the
entanglement dynamics is  significantly different.
Sec.~\ref{sec:3su1g0g} shows an example in which three electronic
states are populated by a sequence of two chirped laser pulses. The
vibrational wave packets excited in each electronic potential are much
more complex, having various localizations and intricate vibrational
motions.

\begin{figure}
\includegraphics[width=0.95\columnwidth]{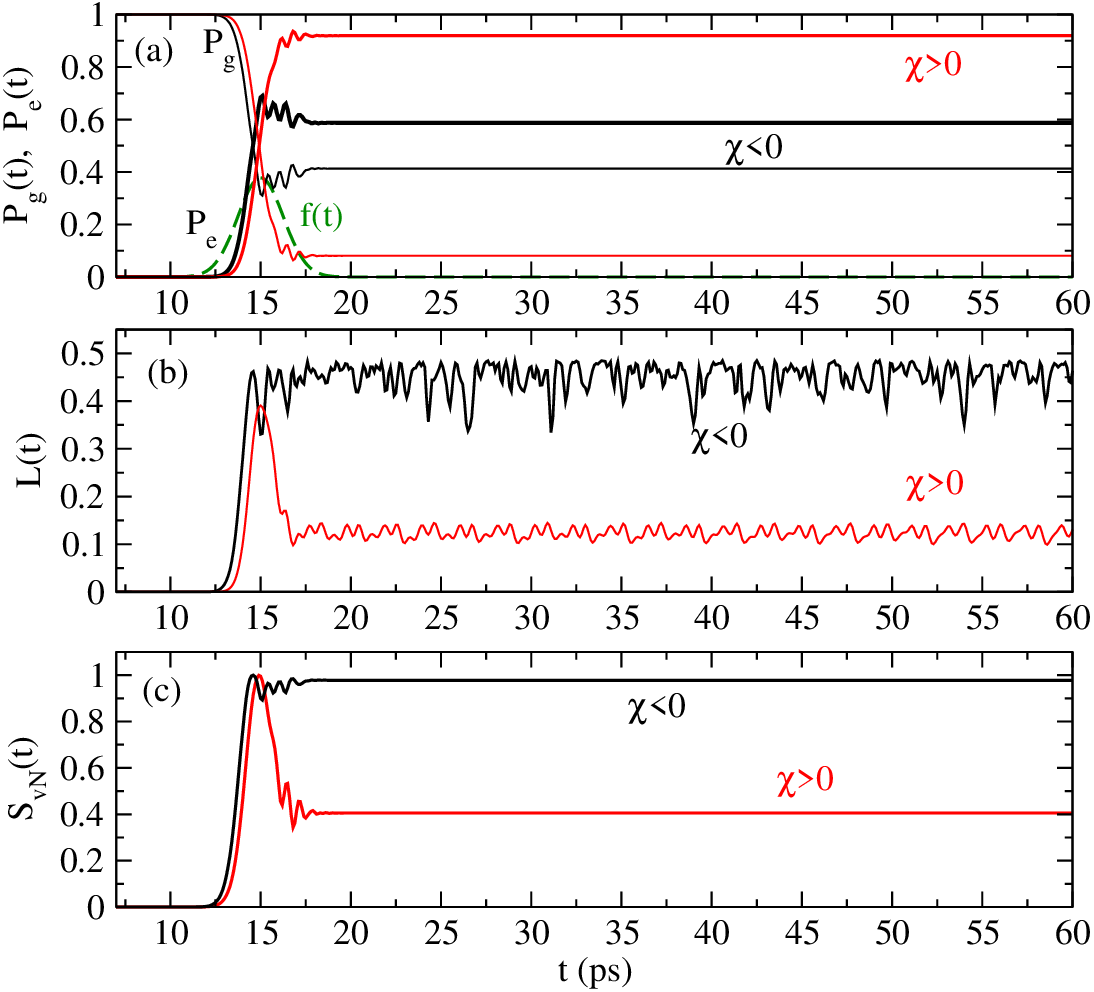}%
 \caption{\label{fig2_3Su1gv0cp} (Color online) Control of the
   electronic-nuclear entanglement dynamics by the sign of the chirp
   rate $\chi$, for a coupling $W_L=26.34$ cm$^{-1}$ between
  the electronic states
 $g=a^3\Sigma_{u}^{+}$ and $e=1_g$ of Cs$_2$ (Fig.~\ref{fig1_3Su1g}).
(a) Time evolution of the
populations $P_g(t)$ and $P_e(t)$ for positive and negative chirp. The
Gaussian pulse envelope $f(t)$, centered at  $t_P=15$ ps and with
temporal width $\tau_C=2.1$ ps, is represented with dashed line.
(b) Time evolution of the linear entropy $L(t)$ for positive and
negative chirp. (c) Time evolution of the von Neumann entropy
$S_{vN}(t)$ for positive and negative chirp.}
\end{figure}

\subsection{\label{sec:3Su1g} Controlling the electronic-nuclear
  entanglement dynamics in a molecule by populating 
two electronic states with a chirped laser pulse.}

We consider the Cs$_2$ molecule in which  the electronic channels $g=a^3\Sigma_u^+(6s,6s)$ and
$e=1_g(6s,6p_{3/2})$  are coupled by a chirped laser pulse
(Fig.~\ref{fig1_3Su1g}),  described by the electric field
\begin{equation}
{\cal E}(t)= {\cal {E}}_0 f(t) \cos [\omega_L t + \varphi(t)],
\end{equation}
 with amplitude ${\cal {E}}_0$ and Gaussian temporal envelope $f(t)$.
 A chirped pulse \cite{cao98,*cao00} is characterized by several parameters belonging to the spectral and temporal
domains, which can be used  to control the system
evolution \cite{chirpPRA04,elianeepjd04,vatasescu12}.  
$\omega_L/2\pi$ is the central frequency of the pulse, reached at $t=t_{P}$,
and $\varphi(t)$  is a phase which is a quadratic function of time,
such that the instantaneous frequency $\omega(t)=\omega_L +d\varphi/dt$ varies linearly 
with the chirp rate $\chi$ around the central frequency $\omega_{L}/2 \pi$:  $\omega(t)=\omega_{L}+ \chi(t-t_P)$.
The Gaussian envelope $f(t)=\sqrt{\tau_L/\tau_C}
\exp \lbrace -2 \ln 2 [(t-t_P)/\tau_C]^2 \rbrace$ is centered at $t=t_{P}$, having the temporal width $\tau_{C}$.
  The duration $\tau_{L}$ is the temporal width of the transform limited
pulse (before chirping),  and characterizes the spectral width of the
pulse in the frequency domain: $\delta \omega =4 \ln 2/\tau_L$.
The chirp rate $\chi$
\footnote{Related to the ratio $\tau_ C / \tau_L \ge 1 $ by
$\tau_ C / \tau_L = \sqrt{ 1 + (\chi^2\tau^4_ C)/ (4 \ln 2)^2}$} 
 and its sign are essential control parameters.   The sign of the chirp 
 determines the sense of sweeping the difference $V_g(R) - V_e(R)$ between the
 electronic potentials, by increasing or decreasing the instantaneous
 frequency of the pulse $\omega(t)$ (see Fig.~\ref{fig1_3Su1g}), which
 leads to the excitation of different vibrational wave packets.

\begin{figure}
\includegraphics[width=0.9\columnwidth]{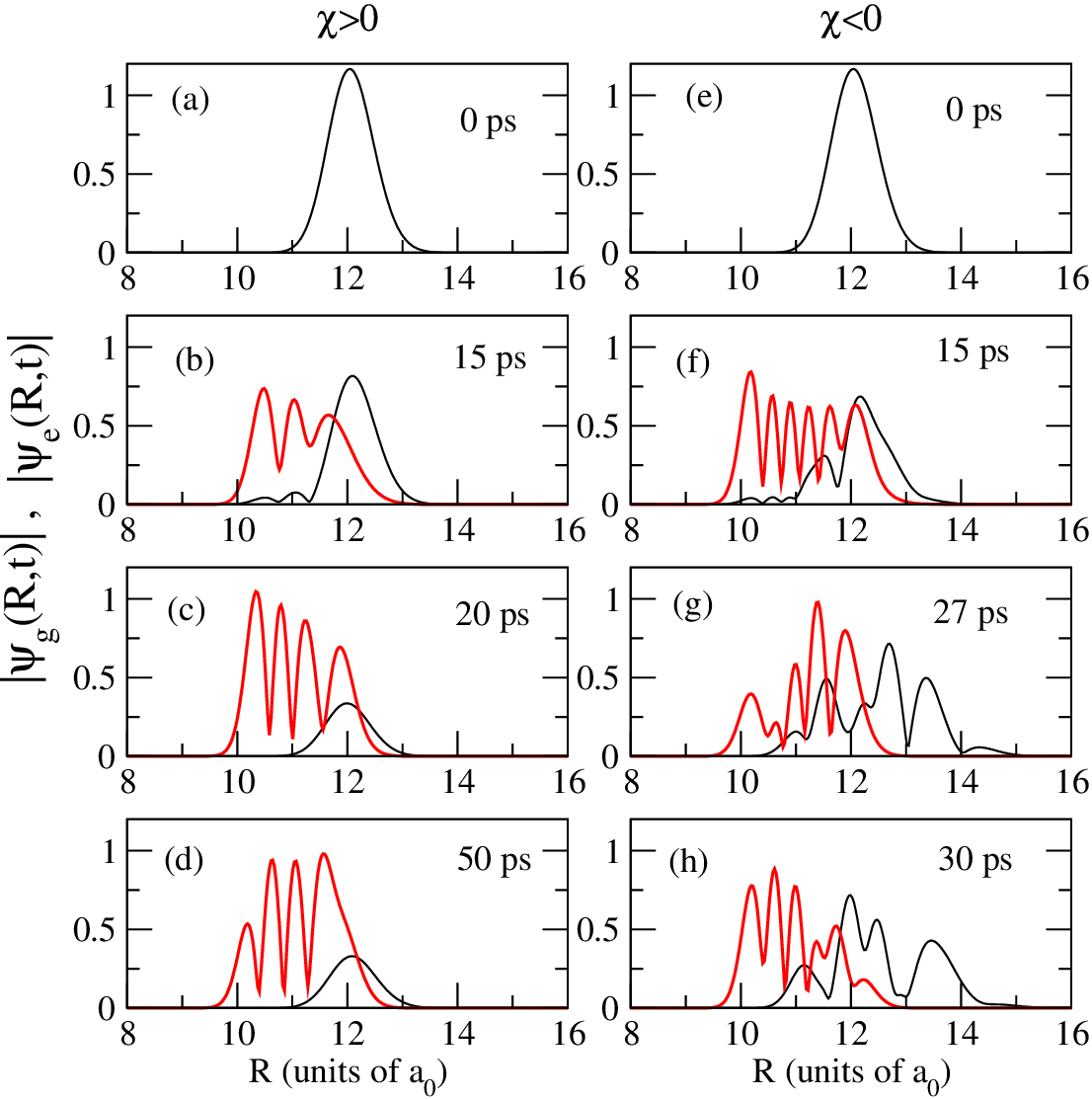}%
\caption{\label{fig3_wfchirp} (Color online) Time evolution of the
  vibrational components $\psi_{g}(R,t)$ (thin line)  and
  $\psi_{e}(R,t)$ (thick line) of the pure
entangled state $|\Psi_{el,vib}(t) >$, created by a chirped pulse. (a-d, left column) Time evolution of
$|\psi_{g}(R,t)|$,  $|\psi_{e}(R,t)|$ for positive chirp, $\chi >0$. (e-h, right column) Time evolution of
$|\psi_{g}(R,t)|$,  $|\psi_{e}(R,t)|$ for negative chirp, $\chi <0$.}
\end{figure}

Here we consider  a chirped pulse with central energy $\hbar \omega_L=$ 10695 cm$^{-1}$ which couples the electronic potentials
$V_g(R)=a^3\Sigma_u$ and $V_e(R)=1_g$ of Cs$_2$ around the
internuclear distance $R_c \approx 12$ a$_0$, transferring
population from the ground state $v_g=0$ of $g=a^3\Sigma^+_u$ to several
low vibrational levels $v_e$ of the excited state $e=1_g$. The
process is represented in Fig.~\ref{fig1_3Su1g}, the 
electronic curves being those described in \cite{vatasescu09}. We
suppose a
chirped pulse with the envelope $f(t)$ centered at  $t_P=15$ ps,
and temporal width $\tau_C=2.1$ ps (represented in
Fig.~\ref{fig2_3Su1gv0cp}(a)),  obtained by chirping a  transform limited
pulse with duration $\tau_L=0.3$ ps (spectral width $\delta \omega =49$
cm$^{-1}$), using a chirp rate $|\chi|=4.35$
ps$^{-2}$. The  energy range  swept by the chirped pulse around the central
frequency $\omega_{L}/2 \pi$ is $2\hbar |\chi| \tau_C$ \cite{elianeepjd04}, with $\hbar
|\chi|=23.11$ cm$^{-1}/$ps, allowing the excitation of several
vibrational levels in the $1_g$ potential, where the vibrational level
spacing in the excitation range is about 16 cm$^{-1}$.

The time-dependent Schr\"odinger equation describing the dynamics of
the vibrational wave packets $\psi_{g,e}(R,t)$ in the electronic channels coupled by the
pulse, written using the rotating wave approximation with the frequency
$\omega_L/2\pi$ \cite{chirpPRA04,vatasescu12},  is
\begin{eqnarray}
\label{eqS2states}
&&i\hbar\frac{\partial}{\partial t}\left(\begin{array}{c}
 \Psi_{e}(R,t)\\
\Psi_{g}(R,t)
 \end{array}\right)=\\
&&
\left(\begin{array}{lc}
 {\bf \hat T} + V^{\prime}_{e}(R)  & W_L f(t) e^{-i \varphi(t)} \\
 W_L f(t) e^{i \varphi(t)}  & {\bf \hat T} + V^{\prime}_g(R)
 \end{array} \right)
 \left( \begin{array}{c}
 \Psi_{e}(R,t)\\
\Psi_{g}(R,t)
 \end{array} \right). \nonumber
\end{eqnarray}
 In Eq.~(\ref{eqS2states}), ${\bf \hat T}$ is the kinetic energy operator, and  $V^{\prime}_{e}(R)=V_{e}(R)$,
 $V^{\prime}_g(R)=V_g(R)+\hbar \omega_L$ are the diabatic
 potentials dressed with the energy $\hbar \omega_L$.  $W_L= {\cal {E}}_0 D_{ge}/2$ is the strength of the laser
 coupling depending on the laser intensity $I$ (${\cal
{E}}_0=\sqrt{2I / c\epsilon_0}$) and on the transition dipole moment
$D_{ge}$ between the electronic surfaces \cite{vatasescu01}. Here we just
use a constant strength coupling $W_L$ to explore time evolution under
various pulse parameters.

\begin{figure}
\includegraphics[width=0.95\columnwidth]{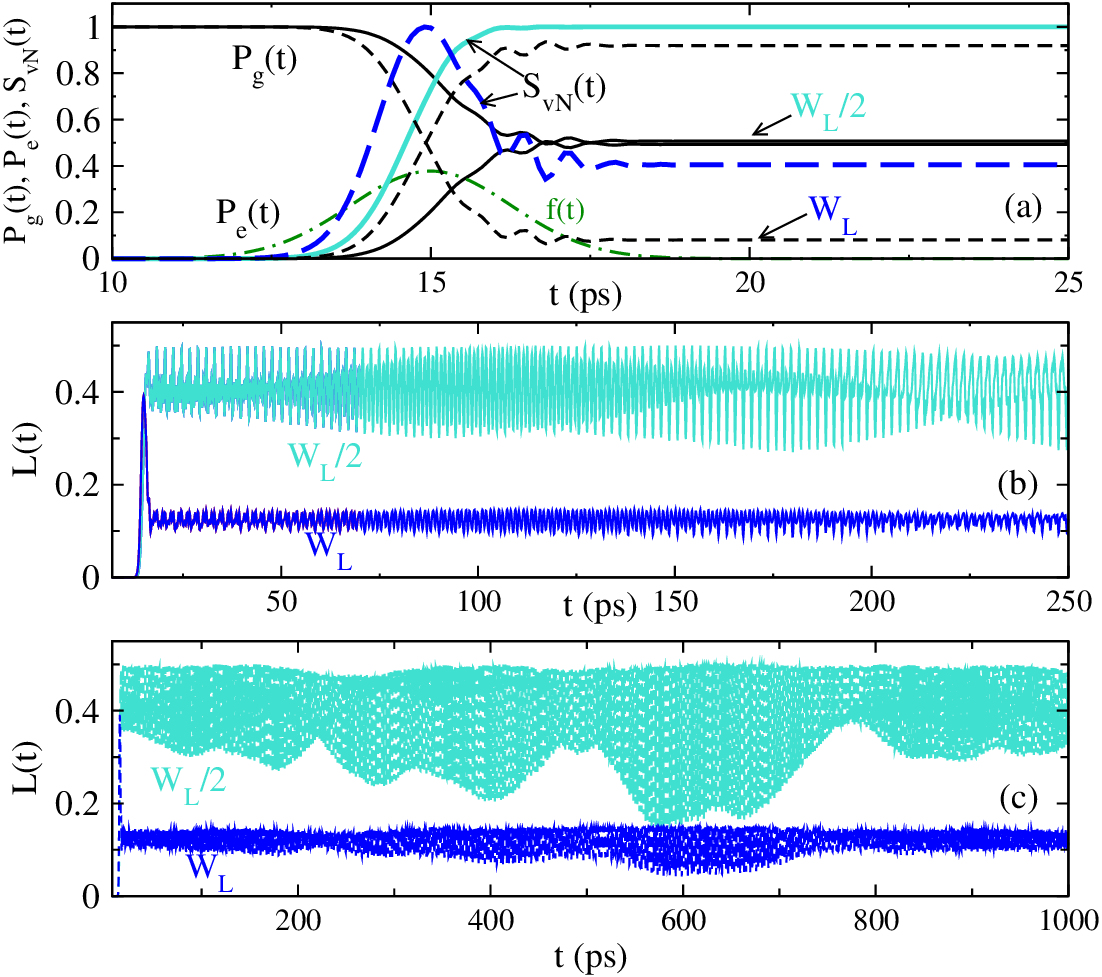}%
 \caption{\label{fig4_3Su1gv0long} (Color online) Control of  electronic-nuclear
   entanglement dynamics in  Cs$_2$ through the
   strength $W_L$ of the chirped  pulse coupling   the electronic states $g=a^3\Sigma_{u}^{+}$ and
   $e=1_g$ (Fig.~\ref{fig1_3Su1g}).
 Results for  $W_L=26.34$ cm$^{-1}$ and
$W_L/2$, the last one producing ``maximum electronic-nuclear entanglement''.
(a) Time evolution of the
populations $P_g(t)$, $P_e(t)$ (full line for $W_L/2$, dashed line for
$W_L$), and of the von Neumann entropy $S_{vN}(t)$ (full line for $W_L/2$, dashed line for
$W_L$) during the pulse. The
Gaussian pulse envelope $f(t)$ is represented with dot-dashed
line. (b,c) Time evolutions of the linear entropy $L(t)$ after pulse:
(b) until  250 ps; (c)  until 1000 ps.}
\end{figure}

 The Schr\"odinger equation (\ref{eqS2states}) is solved numerically 
 by propagating
in time the initial wavefunction
$\left(\begin{array}{c}
0 \\
\chi_{v_g=0}(R)
\end{array}\right)$
on a spatial grid  with length $L_R$, $\chi_{v_g=0}(R)$ being
the vibrational eigenstate with $v_g=0$ in the $a^3\Sigma^+_u$
potential,  represented in Fig.~\ref{fig1_3Su1g} and in Figs.~\ref{fig3_wfchirp}(a),(e).
 The time propagation uses the Chebychev expansion
of the evolution operator  \cite{kosloff94,kosloff96} and the Mapped Sine Grid (MSG) method \cite{elianeepjd04,
 willner04}  to represent the radial dependence of the wave packets. The populations in each electronic state
are calculated from the vibrational wave packets $\Psi_{g,e}(R,t)$ as $P_{g,e}(t) = \int_0^{L_R}
| \Psi_{g,e}(R',t) |^{2} dR'$, with the total population normalized at 1 on the spatial grid ($P_g(t)+P_e(t)=1$),
and $P_g(0)=1$. The von Neumann entropy $S_{vN}(t)$ and the linear
entropy $L(t)$ are calculated using the formulas (\ref{vonNel})  and
(\ref{linentr2}).

\begin{figure}
\includegraphics[width=0.8\columnwidth]{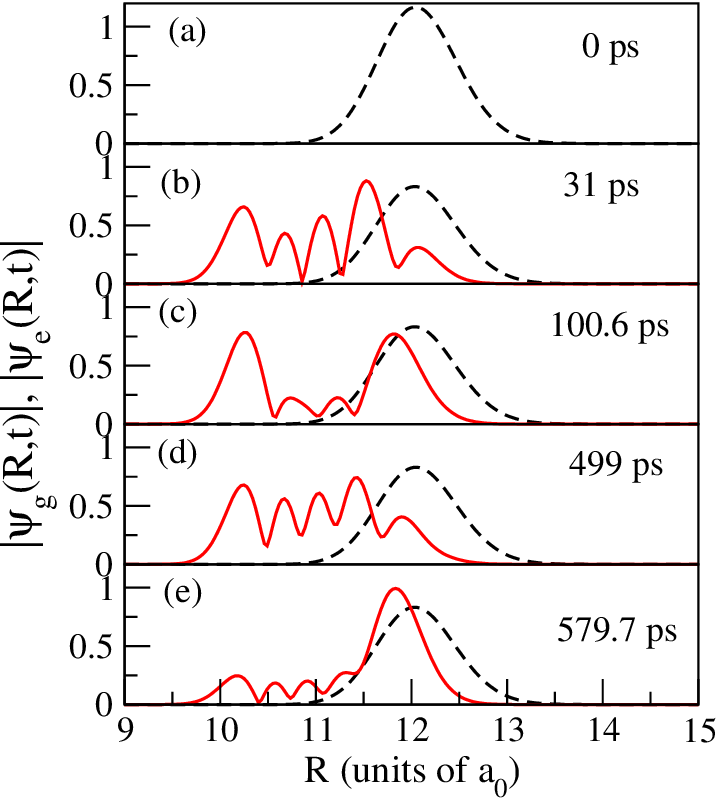}%
\caption{\label{fig5_wflong} (Color online)  Time evolution of the
  vibrational wave packets
$|\psi_{g}(R,t)|$ (dashed line) and $|\psi_{e}(R,t)|$ (thick line) for
the case of \text{``}maximum entanglement\text{``} achieved by the
chirped pulse for a coupling $W_L/2$  (Fig.~\ref{fig4_3Su1gv0long}).}
\end{figure}

Figs.~\ref{fig2_3Su1gv0cp},\ref{fig3_wfchirp} show results
obtained  for a positive or a  negative chirp rate  $\chi$,
for the same coupling  $W_L=26.34$ cm$^{-1}$.  We see that, by
changing the chirp sign,  significantly different results are obtained. The pulse with positive chirp $\chi >0$ begins excitation
from the lowest $v_e$ levels in $1_g$,  producing an inversion of
population between the two electronic channels
(Fig.~\ref{fig2_3Su1gv0cp}(a)) and a \text{``}small\text{``}
entanglement: the von Neumann entropy after pulse is 
$S_{vN}(t)=0.4$ (Fig.~\ref{fig2_3Su1gv0cp}(c)) and the linear
entropy oscillates around 0.1 (Fig.~\ref{fig2_3Su1gv0cp}(b)).
The time evolution of the
wave packets is shown in Figs.~\ref{fig3_wfchirp}(a-d). In the
electronic state
$g=a^3\Sigma^+_u$ the fundamental vibrational state $v_g=0$ (which is
the initial state of the process) is the
only one populated. The pulse populates the vibrational levels with
$v_e=2,3$ in  the excited state $1_g$,
separated by $\approx$ 16 cm$^{-1}$, which is reflected in the oscillations of
about 2 ps in the linear entropy after pulse
(Fig.~\ref{fig2_3Su1gv0cp}(b)).  Indeed,  in
Sec.~\ref{sec:linentrosc} we have shown that this is the
characteristic time to be expected in the
linear entropy evolution  in a $2 \times 3$ system (one level $v_g$ populated in $g$ electronic state, and two levels
$v_e, v'_e$ in $e$ electronic state), and it coincides with the
vibrational period $T_{vib}(v_e=3)=2$ ps.

On the contrary, if the chirp is
negative, $\chi <0$, the pulse begins by exciting higher vibrational
levels in $1_g$, and continues with  lower vibrational levels. A superposition of
vibrational states dominated by $v_e=4,5$ is excited in $1_g$, and also a
superposition of vibrational levels (mainly $v_g=3,4,5$) remains populated in $a^3\Sigma^+_u$ (Figs.~\ref{fig3_wfchirp}(e-h)).
This gives a stronger entanglement: the von Neumann entropy after pulse
is close to 1 (Fig.~\ref{fig2_3Su1gv0cp}(c)).  After pulse, the
linear entropy
(Fig.~\ref{fig2_3Su1gv0cp}(b))
is a highly oscillating function, whose amplitude varies between 0.33
and 0.5. Since several vibrational states are populated in each
electronic potential, there are several characteristic times $T_{osc}$ 
interwined in $L(t)$ evolution, according to the analysis made in Sec.~\ref{sec:linentrosc}.

We shall consider now the formation of an entangled state
$|\Psi_{el,vib}(t) >$ using the coupling strength
$W_L$ as a control parameter.
Fig.~\ref{fig4_3Su1gv0long} shows results obtained with a chirped pulse having the same
parameters as before and positive chirp rate $\chi=4.35$ ps$^{-2}$, for 
the coupling strengths   $W_L=26.34$ cm$^{-1}$ and
$W_L/2$. The case $W_L$ with positive chirp was already analyzed. 
If the coupling is diminished at $W_L/2$, the pulse achieves the
equalization of electronic populations $P_g(t)=P_e(t)=1/2$ (Fig.~\ref{fig4_3Su1gv0long}(a)), creating  
maximum entanglement ($S_{vN}(t)=1$) at the end. The time evolution of the
wave packets is shown in Fig.~\ref{fig5_wflong}, illustrating several
instants of the vibrational motion in the excited electronic state. In the
electronic state
$g=a^3\Sigma^+_u$ only the fundamental vibrational state $v_g=0$ is
populated, and the vibrational superposition  in the excited
state $e=1_g$ is made mainly by the vibrational levels  $v_e=3,4$.
After pulse, the linear entropy is an oscillating function
(Fig.~\ref{fig4_3Su1gv0long}(b)) with the main
oscillation period equal to $T_{vib}(v_e=3)=2$ ps. The long term evolution (until 1000 ps)  
 shows the large amplitude of  the linear entropy variations: 
 $L(t)$ oscillates from a maximum of 0.5 to a minimum of 0.15
 (Fig.~\ref{fig4_3Su1gv0long}(c)).
This large difference between $L(t)$ minima and maxima is due to
the maximization and minimization  of the overlap integral, created by
the vibrational motion of the excited wave packet. 
Figs.~\ref{fig5_wflong}(d,e)  show the vibrational wave packets at $t=499$ ps,
when entanglement is maximal  ($L(t) \approx 0.5$) and the overlap is minimal, and at $t=579.7$
ps, when the entanglement becomes minimal ($L(t) \approx 0.15$)
because the overlap is maximal.

\subsection{\label{sec:3su1g0g} Entanglement dynamics in a case of three electronic potentials coupled by two chirped laser pulses}

\begin{figure} 
\includegraphics[width=0.9\columnwidth]{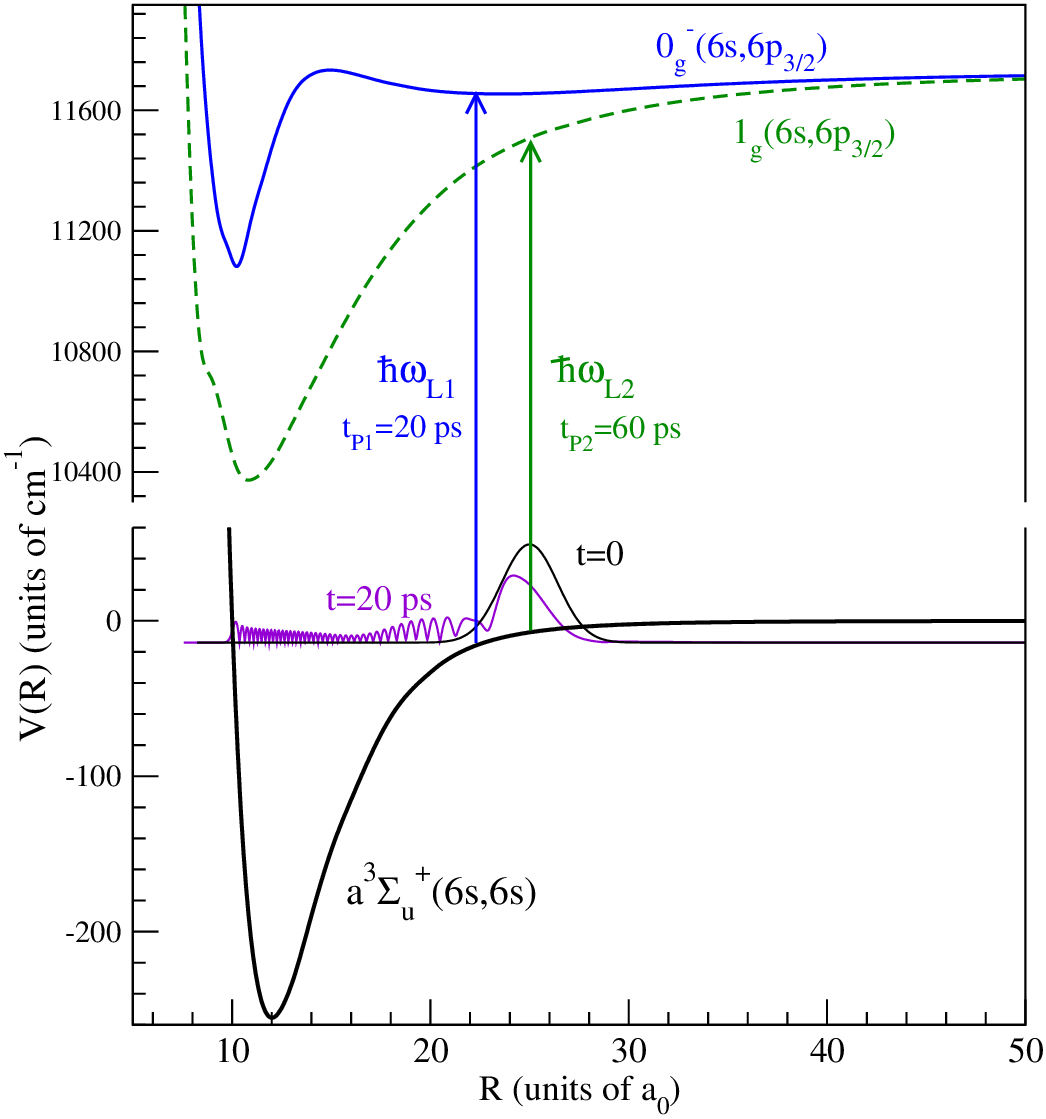}%
 \caption{\label{fig6_3pot} (Color online) $a^3\Sigma_{u}^{+}(6s,6s)$,
  $1_g(6s,6p_{3/2})$, and  $0_g^-(6s,6p_{3/2})$ electronic
potentials of Cs$_2$, coupled by two successive chirped laser
pulses. The first pulse, with central
energy $\hbar \omega_{L1}=11680$  cm$^{-1}$, and $t_{P1}=20$ ps, transfers population from
$a^3\Sigma_{u}^{+}$
to the double well potential $0_g^-(6s,6p_{3/2})$. The second one,
with $\hbar \omega_{L2}=11513$  cm$^{-1}$ and centered at $t_{P2}=60$ ps,  transfers
population from
$a^3\Sigma_{u}^{+}$ to $1_g(6s,6p_{3/2})$. 
The initial state of the process is a Gaussian wave packet in the $a^3\Sigma_{u}^{+}(6s,6s)$
electronic state, represented in the figure. After pulses, all three  electronic
potentials remain populated. The energy origin is taken to be the
dissociation limit $E_{6s+6s}=0$ of the $a^3\Sigma_{u}^{+}(6s,6s)$
potential. }
\end{figure}

\begin{figure}
\includegraphics[width=0.95\columnwidth]{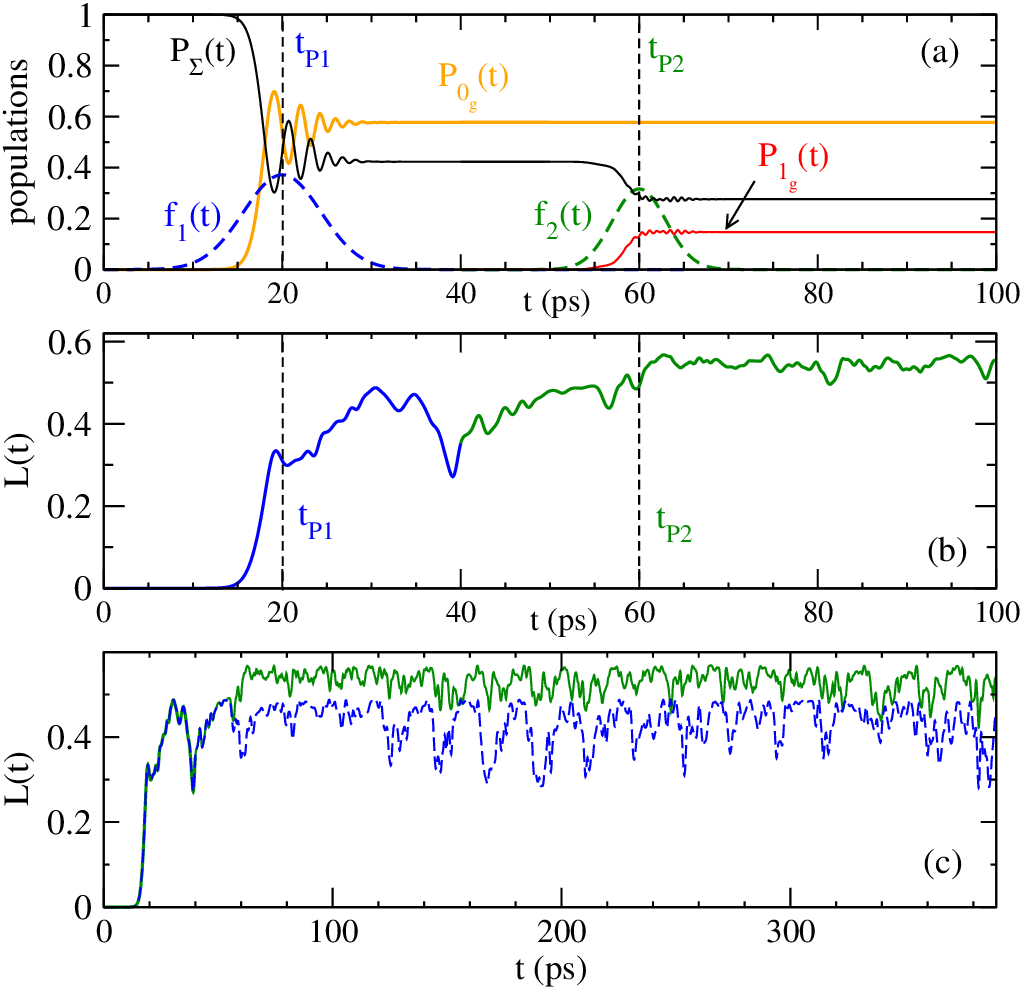}%
 \caption{\label{fig7-linentr3pot} (Color online) 
   Electronic-nuclear entanglement dynamics in the Cs$_2$ molecule, 
created by the sequence of two pulses which populate the three electronic potentials represented in
  Fig.~\ref{fig6_3pot}.
 (a) Time evolution of the populations $P_{\Sigma}(t)$, $P_{0_g}(t)$,
 and  $P_{1_g}(t)$ due to the chirped pulses whose envelopes $f_1(t)$
 and  $f_2(t)$ are represented with dashed line. 
(b) Time evolution of the linear entropy $L(t)$ during the first pulse
(after which two electronic states are populated) and the
second pulse (which populates also the third one). (c) Long term
evolution of the linear entropy $L(t)$.
With dashed line is represented the long term evolution of $L(t)$ in the hypothetical case of the first
pulse only.}
\end{figure}

Let us now consider the Cs$_2$ molecule, in which 
an entangled state $|\psi_{el,vib}(t)>$ is created by a sequence of two chirped laser
pulses, which couple consecutively  the electronic
state $a^3\Sigma_{u}^{+}(6s,6s)$ to $0_g^-(6s,6p_{3/2})$ and to $1_g(6s,6p_{3/2})$.
 The scheme is shown in
Fig.~\ref{fig6_3pot}. The first pulse couples $a^3\Sigma_{u}^{+}$ to
$0_g^-$, leaving both states populated. After the end of
the first pulse, the second pulse couples $a^3\Sigma_{u}^{+}$ to
$1_g$. At the end of the sequence, all three electronic states rest
populated, in a process which increases progressively the entanglement
(from two to three electronic states).

Let us  detail the scheme. The initial state of the process, 
represented in  Fig.~\ref{fig6_3pot}, is a Gaussian
wave packet in the electronic  $a^3\Sigma_{u}^{+}(6s,6s)$ potential,
localized around 25 a$_0$ and simulating a superposition of
vibrational states of $a^3\Sigma_{u}^{+}(6s,6s)$ centered around the state with
$v_{_\Sigma}=36$, which is bounded by $E_{v_{_\Sigma}=36} \approx -17$ cm$^{-1}$. The two chirped pulses have Gaussian temporal envelopes
$f_1(t)$ and $f_2(t)$, which are centered at $t_{P1}=20$ ps and $t_{P2}=60$ ps,
respectively (represented in Fig.~\ref{fig7-linentr3pot}(a)).

\begin{figure}
\includegraphics[width=0.95\columnwidth]{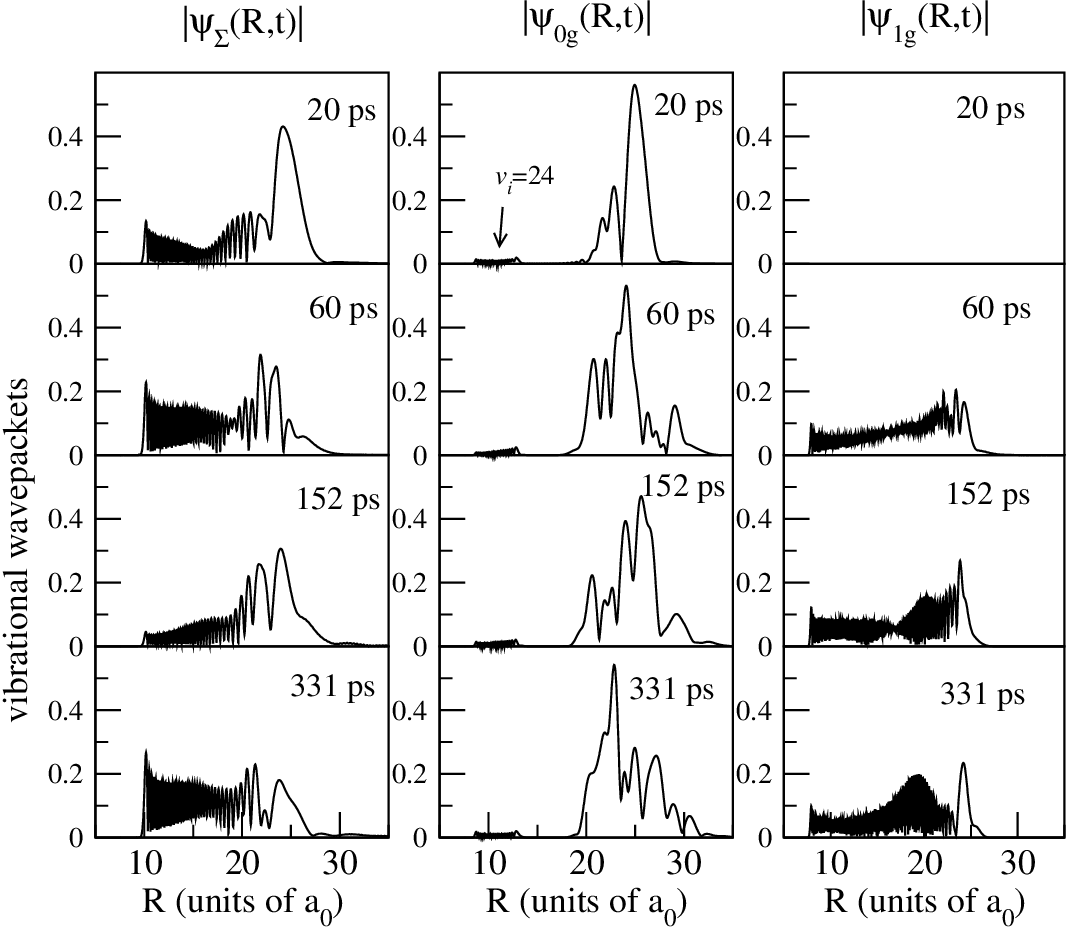}%
\caption{\label{fig8-wf3} Time evolution of the
  vibrational wave packets  $\psi_{_\Sigma}(R,t)$,
  $\psi_{0_g}(R,t)$ and  $\psi_{1_g}(R,t)$ excited by the sequence of
  two chirped pulses in the electronic potentials
  $a^3\Sigma_{u}^{+}(6s,6s)$, $0_g^-(6s,6p_{3/2})$, and
  $1_g(6s,6p_{3/2})$, represented in Fig.~\ref{fig6_3pot}.}
\end{figure}

The first chirped pulse, with central
energy $\hbar \omega_{L1}=11680$  cm$^{-1}$, couples the $a^3\Sigma_{u}^{+}$ electronic state
to the $0_g^-(6s,6p_{3/2})$ state. The pulse has  the temporal width
$\tau_{C1}=7.2$ ps (with $\tau_{L1}=1$ ps) and a positive chirp rate
$\chi_1=0.379$ ps$^{-2}$, such as the energy range resonantly swept around
the central frequency is $2 \hbar |\chi_1|
\tau_{C1} \approx 28$ cm$^{-1}$. The
coupling strength is $W_{L1}=6.6$ cm$^{-1}$. The
first pulse populates a superposition of vibrational levels in the
external well of the $0_g^-(6s,6p_{3/2})$ potential, exciting also
 the vibrational level $v_i=24$ of the $0_g^-$ inner
well. Fig.~\ref{fig8-wf3} shows the vibrational
wave packets $a^3\Sigma_{u}^{+}$ and $0_g^-$ populated by the first
pulse  at t=20 ps. The wave packets evolution during the pulse is obtained by
solving numerically a temporal Schr\"odinger equation similar with
Eq.~(\ref{eqS2states}).   The time evolution of the populations is represented in
Fig.~\ref{fig7-linentr3pot}(a). 

The second pulse, with $\hbar \omega_{L2}=11513$  cm$^{-1}$ and centered at $t_{P2}=60$ ps,  transfers
population from $a^3\Sigma_{u}^{+}$ to $1_g(6s,6p_{3/2})$. The pulse
has a  coupling strength $W_{L2}=26.3$ cm$^{-1}$, temporal width
$\tau_{C2}=5$ ps (with $\tau_{L1}=0.5$ ps) and a positive chirp rate
$\chi_2=1.1$ ps$^{-2}$. The energy range  resonantly swept around
its central frequency  $\omega_{L2}/2 \pi$  is $2 \hbar |\chi_2|
\tau_{C2} \approx 58.6$ cm$^{-1}$, and  a
superposition of high excited vibrational levels (around the level with
$v_{1_g}=108$) is populated in the $1_g$ electronic potential. 

Fig.~\ref{fig8-wf3} shows the dynamics of the vibrational wave
packets in the three
electronic potentials. The time evolution of the electronic
populations is represented in Fig.~\ref{fig7-linentr3pot}(a). The
chirped Rabi periods characteristic for the action of a chirped pulse \cite{chirpPRA04}
are visible during each pulse. 

The linear entropy of entanglement $L(t)$ is calculated using the formula
(\ref{linentrgen}), and its time evolution during the pulse sequence is represented in
Fig.~\ref{fig7-linentr3pot}(b).  By populating a third electronic
state, the second pulse increases the molecular entanglement, as we
have shown in Sec.~\ref{sec:linentrvibcoh}. The long term linear entropy
evolution, after the end of the pulse sequence, is shown in Fig.~\ref{fig7-linentr3pot}(c).
In the same figure we have represented 
$L(t)$  evolution  supposing that only the first pulse would act on
the molecule, and therefore only  two
electronic states would be populated. In this case the entanglement
dynamics is due to vibronic coherences between only two electronic
states, showing large variations between minima and maxima. As we have
shown in Sec.~\ref{sec:CohResource}, this large amplitude in $L(t)$
variations is an indicator for  the strength of the electronic
coherence measured by $C_{l_1}( \hat{\rho}_{el})$, which is proportional
to the overlap $|<\psi_{g}(R,t)|\psi_{e}(R,t)>|$. 
When three electronic states are populated, entanglement is increased and
$L(t)$ variations in time are diminished. This shows a decreasing of 
the electronic coherence measured by $C_{l_1}( \hat{\rho}_{el})$,
 due to smaller overlaps between the three vibrational wave packets.

Therefore,  we have shown  examples of a molecule  prepared in an electronic-vibrational
entangled state by chirped laser pulses which create  coherent vibrational
wave packets in several electronic potentials. Dephasing and
recurrence due to periodic oscillations are specific to wave packets
vibrational motion in bound electronic potentials.
Electronic-nuclear entanglement
oscillations in an isolated molecule so prepared with laser pulses
are indicative for phenomena of
electronic coherence in the molecular system and  periodicity specific to
vibrational motions \cite{zewail93,*aspuru12,*cina12}. Entanglement
may be increased by increasing the number of  populated electronic
states. On the other hand, entanglement oscillations, expressed in the
  temporal variations of the linear entropy, may 
be of  large amplitude, and can be controlled by  quantum preparations.

\section{\label{conclu} Conclusion}

We have derived measures of entanglement and quantum coherence for a 
molecular system described in a bipartite Hilbert space
$\cal{H}$$=$$\cal{H}$$_{el}$$\bigotimes$$\cal{H}$$_{vib}$ of dimension
$N_{el} \times N_v$, establishing relations
 between the linear entropy of electronic-vibrational
entanglement and quantifiers of quantum coherence in the bipartite molecular system.

For a Hilbert space of dimension $2 \times N_v$, we have discussed 
the expressions for  the von Neumann and  linear entropy of
electronic-nuclear entanglement \cite{vatasescu2013},  showing that a remarkable difference
between these two measures of entanglement  appears when their
temporal behaviours in the case of an isolated molecule are considered.
In contrast to the
von Neumann entropy of entanglement,  the linear entropy
\text{"}understands\text{"} vibrational motion in the electronic potentials
as entanglement dynamics.
We find linear
entropy of entanglement as being a more complex informational quantity,
recalling  previous assertions  about the \text{"}conceptual inadequacy\text{"} \cite{zeilinger01}
of the von Neumann entropy in defining the
 information content of a quantum system.  These discussions were accompanied by proposals for a
more appropriate measure, which, interestingly,  has  proven to be
essentially the linear entropy \cite{zeilinger99,zeilinger01,luo06}.

We have derived the linear entropy of electronic-vibrational entanglement for a bipartite
Hilbert space $\cal{H}$$=$$\cal{H}$$_{el}$$\bigotimes$$\cal{H}$$_{vib}$ 
with dimension $N_{el} \times N_v$, showing its
dependence on the vibronic coherences of the molecule, a 
 property that connects this entanglement measure  to coherence quantifiers.

Quantum coherence in the bipartite entangled
state $\hat{\rho}_{el,vib}(t)$ was characterized employing the
resource approach \cite{baumgratz14,girolami14}, using measures of coherence based
on  $l_1$ norm and Wigner-Yanase skew information. Connections between
quantum coherence, quantum uncertainty in energy, and the \text{"}velocity\text{"}
 of $\hat{\rho}_{el,vib}(t)$ evolution \cite{anandan90} are outlined
 in Sec.~\ref{sec:CohUncert}. 

We have employed the skew information as a measure of quantum
coherence and quantum uncertainty in the pure entangled state
$\hat{\rho}_{el,vib}(t)$  and in the reduced electronic state 
 $\hat{\rho}_{el}$,  taking as observables  the Hamiltonians
 $\hat{H}_{mol}$ and $\hat{H}_{el}$. 
 We have derived the  Wigner-Yanase skew information in the reduced electronic state 
 $\hat{\rho}_{el}$ for the electronic Hamiltonian $\hat{H}_{el}$, 
and  in the pure entangled state
$\hat{\rho}_{el,vib}(t)$ for the observables $\hat{H}_{mol}$
(molecular Hamiltonian) and
$\hat{H}_{el}\bigotimes \hat{I}_v$ (local observable $\hat{H}_{el}$), 
for a bipartite Hilbert space of dimension $N_{el} \times N_v$.
We have shown that linear entropy of entanglement is connected to the skew
information ${\cal I_S}(\hat{\rho}_{el,vib}, \hat{H}_{el}\bigotimes
\hat{I}_v)$ and  ${\cal I_S} (\hat{\rho}_{el}, \hat{H}_{el})$, related to the measurement of the local observable
$\hat{H}_{el}$ in the correlated quantum systems  (el$\bigotimes$vib).

The characteristic times of entanglement dynamics due to vibrational
motion in the electronic potentials  are analyzed in
Sec.~\ref{sec:linentrosc}.
In the last part of this paper, Sec.~\ref{sec:3Su1g},  we show
examples  of  these entanglement
oscillations for the  $Cs_2$ molecule prepared in an electronic-vibrational
entangled state by chirped laser pulses which create  coherent vibrational
wave packets in several electronic potentials.    We have shown the control of  entanglement dynamics
by using chirped laser pulses, whose parameters can be chosen to
create specific quantum preparations 
and significant changes in  entanglement dynamics.
 
We hope that the present work will contribute to the ample research
program intended to enlighten our understanding of molecular
phenomena by using quantum information concepts.

\begin{acknowledgments}
This work was supported by the LAPLAS 3 39N Research Program of the Romanian Ministry of Education
and Research.
\end{acknowledgments}

\bibliography{art-entcoh}% Produces the bibliography via BibTeX.

\end{document}